%% file: main.tex
\documentclass{article}
\usepackage{graphicx} 
\usepackage{amsmath}
\usepackage{amsthm}
\usepackage{amsfonts}
\usepackage{dsfont}
\usepackage{xcolor}
\usepackage{enumitem}
\usepackage{tabto}
\usepackage{tikz}
\usetikzlibrary{patterns,snakes}
\usepackage{subcaption}
\usepackage{comment}
\usepackage{biblatex}
\usepackage{booktabs}
\usepackage{verbatim}
\usepackage{multirow}
\usepackage{makecell}
\usepackage{float}
\usepackage{longtable}
\usepackage{siunitx}
\newtheorem*{definition*}{Definition}

\usepackage[a4paper, total={7in, 10in}]{geometry}

\definecolor{crimson}{RGB}{221, 19, 58}
\definecolor{navy}{RGB}{0, 0, 128}
\definecolor{forestgreen}{RGB}{36, 139, 33}

\definecolor{indigo}{RGB}{76, 0, 130}
\definecolor{lightseagreen}{RGB}{33, 176, 171}
\definecolor{gold}{RGB}{215, 167, 32}

\title{MARS: A framework for modelling register-based social networks}
\author{Katherine Hamilton\textsuperscript{1}, Irina Epure\textsuperscript{1}, Frank Takes\textsuperscript{1,*}}
\date{%
    \small \textsuperscript{1}Leiden University\\%
    \textsuperscript{*}f.w.takes@liacs.leidenuniv.nl%
}

\newcommand{\al}[1]{a_{#1}^\ell}
\newcommand{\Al}[1]{A_{#1}^\ell}

\newcommand{\prob}[1]{\mathbb P \left(#1\right)}
\newcommand{\expected}[1]{\mathbb E \left[#1\right]}

\newcommand{\dx}[1]{\text{ d}#1}

\newcommand{\ifc}{\text{ if }}

\newcommand{\given}{\mid}

\newcommand{\pdf}[1]{f_{#1}}

\newcommand{\Gl}{G^\ell}
\newcommand{\Hl}{H^\ell}
\newcommand{\vl}[1]{#1^\ell}
\newcommand{\Vl}{V^\ell}
\newcommand{\El}{E^\ell}

\newcommand{\XV}{\mathbf X_V}
\newcommand{\xu}{\mathbf x_u}
\newcommand{\xv}{\mathbf x_v}
\newcommand{\XA}{\mathbf X_{\mathcal A}}
\newcommand{\xal}[1]{\mathbf x_{\al{#1}}}
\newcommand{\w}[2]{\omega\big(d(#1, #2)\big)}

\newcommand{\alset}[1]{\{\al{#1}\}_{#1=1}^K}
\newcommand{\xalset}[1]{\{\mathbf x_{\al{#1}}\}_{#1=1}^K}

\newcommand{\posR}{\mathbb R^+}
\newcommand{\sqR}{\mathbb R^2}

\newcommand{\inl}{\sim_{\ell}}

\newcommand{\cli}[4]{c_{#2}^{#1}(#3, #4)}

\begin{document}

\maketitle

\begin{abstract}
    Register-based social networks have become of increasing interest in countries where formal government-curated microdata is available. 
    Due to the non-trivial generative process of register-based networks, existing random graph models fail to facilitate effective structural analysis, hindering the discovery of meaningful insights in the underlying social system.
    In this paper we introduce the Multiplex Affiliation-based Random Spatially-embedded (MARS) graph framework, which replicates the construction method of register-based social networks. 
    We derive fundamental statistical properties of MARS ensembles in general and special cases. 
    To demonstrate the applicability of the framework, we implement a simple model under the MARS framework and show that it recovers similar properties to those exhibited by the population-scale register-based social network of the Netherlands. Furthermore, we analyse the effect of spatial tie strength on closure in the network and compare our results with existing empirical findings, showing that increased spatial freedom is correlated with decreased social cohesion. 
\end{abstract}

\section{Introduction}
\label{sec:introduction}
\input{section/1-introduction/introduction}

\section{Preliminaries}
\label{sec:preliminaries}
\input{section/2-preliminaries/preliminaries}

\section{Related Work}
\label{sec:related-work}
\input{section/3-related-work/related-work}

\section{MARS Network Model}
\label{sec:proposed-approach}
\input{section/4-proposed-approach/proposed-approach}

\section{Statistical properties}
\label{sec:statistical-properties}
\input{section/5-statistical-properties/statistical-properties}

\newpage
\section{Experiments}
\label{sec:experiments}
\input{section/6-experiments/experiments}

\section{Conclusion}
\label{sec:conclusion}
\input{section/7-conclusion/conclusion}

\newpage
\printbibliography

\newpage
\section*{Acknowledgements}
The authors thank the members of PLANET-NL for their helpful conversations and suggestions.
\input{section/acknowledgments}

\newpage
\appendix
\section{Expected number of three-dimensional triangles and two-dimensional wedges in a space-independent MARS network}
\label{app:triangle-counting}
\input{appendix/three-dimensional-triangles-space-independent-case}

\input{appendix/two-dimensional-wedges-space-independent-case}

\section{Degree statistics}
\label{app:fitting}
\setcounter{table}{0}
\renewcommand{\thetable}{B\arabic{table}}
\input{appendix/fitting-tables}

\section{Properties of the resulting MARS models}
\label{app:properties-table}
\setcounter{table}{0}
\renewcommand{\thetable}{C\arabic{table}}
\input{appendix/property-tables}

\end{document}

%% file: section/1-introduction/introduction.tex

Random graph models for social network data~\cite{social-network-review} are useful for understanding the mechanisms that govern tie formation and in analysing dynamic processes on network structures \cite{social-network-comparison}. Most random graph models are built from simple connection functions based on pairwise relationships between nodes. However, many real-world social network structures have more complex edge formation processes, and thus require more specialised models to capture their properties. 
An example of more complex network structures which have attracted considerable interest in recent years are social networks constructed from government-curated register data, also referred to as population-scale social networks~\cite{netherlands-anatomy, swedish-anatomy}, nation-scale social networks~\cite{danish-anatomy} or register-based social networks~\cite{menyhert2025connectivity}. Compared to traditional online, survey-based, digital-trace or co-occurrence networks, register-based networks capture more formal ties~\cite{takes2026population} and represent contact opportunity structure, offering different insights into the structure and dynamics of social systems \cite{netherlands-anatomy}.  

In these networks, pairwise relationships are inferred by national statistics bureaus \cite{popnet-data} from government registers or tax filings, capturing work, school, family, home and neighbourhood ties. The resulting networks are: (i)~\textit{multiplex}, as edges are inferred from different registers representing relationships in different social contexts; (ii)~\textit{spatially-embedded}, where individuals have a home address and connections are inferred based on proximity, and; (iii)~\textit{affiliation-based}, as connections are based on co-affiliation to workplaces, schools and addresses. 

The key challenge addressed in this paper is that the method of inferring connections from register data inevitably introduces artificial properties to the network. This means that if one wishes to validate empirical findings related to network structure one must use an appropriate network model which sufficiently reproduces this non-trivial generative process. Using such a model allows us to  distinguish between properties which are truly reflective of the underlying
social system and those which are consequences of the network construction process. 


In this paper, we propose a new Multiplex, Affiliation-based, Random, Spatially-embedded network model (MARS) which captures these fundamental properties of register-based social networks. We develop a flexible and highly-customisable software framework to generate MARS networks, which we employ in a set of empirical experiments. Our first contribution is an exploration of the model's fundamental summary statistics in general and limiting cases. We then demonstrate the applicability of the model to the creation of register-based social networks, by implementing a simple instance of a MARS model based on a soft random geometric connection function \cite{waxman} which generates a network ensemble with similar properties to the well-studied register-based network of the Netherlands \cite{netherlands-anatomy}. Thirdly and finally, we analyse the effect of model parameters on the network structure. We find that when spatial proximity strongly determines edge formation, network closure increases, which agrees with the existing empirical findings of \textcite{fragmentation} that increased mobility leads to
decreased social cohesion. 


The remainder of this article is structured as follows. First, we establish notation and definitions in Section \ref{sec:preliminaries}, and review existing random network models and their limitations in the context of register-based social networks in Section \ref{sec:related-work}. Then we introduce the MARS model for multiplex, spatially-embedded, affiliation-based network data (Section \ref{sec:proposed-approach}) and analyse its basic statistical properties (Section \ref{sec:statistical-properties}). We simulate a soft random geometric MARS ensemble and analyse the properties of the model in the context of register-based social networks in Section \ref{sec:experiments}, and conclude our findings in Section \ref{sec:conclusion}.

%% file: section/2-preliminaries/preliminaries.tex
In this section we review definitions and introduce notation for multiplex, spatially-embedded and affiliation-based networks. 

\subsection{Multiplex networks}
\label{sec:multiplex-networks}
\input{section/2-preliminaries/multi-layer-networks}

\subsection{Spatially-embedded networks}
\input{section/2-preliminaries/spatial-networks}

\subsection{Affiliation-based networks}
\input{section/2-preliminaries/affiliation-based-networks}

%% file: section/2-preliminaries/multi-layer-networks.tex
 \textit{Multiplex networks} are a subclass of multi-layer networks in which each layer is defined on (a subset of) the same set of nodes, and inter-layer edges exist only between two instances of the same node in different layers. In the notation of \textcite{multilayer-networks} we define a multiplex network on $N$ nodes and $L$ layers as a sequence of graphs $\{\Gl\}_{\ell=1}^L$ where each graph $\Gl$ is an ordered tuple $(\Vl, \El)$: $\Vl$ is the node set $\{\vl{u} \given u \in \{1, \dots N \}\}$ and $\El \subseteq \Vl \times \Vl$ is the set of intra-layer edges $(\vl{u}, \vl{v})$ in layer $\ell$. 
 Here $u$ is used to denote the set of instances of the same vertex in each layer $\{\vl{u}\}_{\ell=1}^L$, but we will commonly refer to $u$ simply as a node. For a pair of nodes $u$ and $v$, $u \inl v$ denotes the property that $u$ is adjacent to $v$ in layer $\ell$, i.e. that $(\vl{u}, \vl{v}) \in E^\ell$. The \textit{monoplex} representation of the network is a graph $G=(V, E)$ where $V$ is the node set $\{u \given u \in \{1, \dots N \}\}$ and an edge $(u, v)$ is in $E \subseteq V \times V$ iff $\exists ~\ell \in \{1, \dots L\}$ such that $u \inl v$.  We denote the property that two nodes $u$ and $v$ are adjacent in the monoplex representation $G$ simply by $u \sim v$.

%% file: section/2-preliminaries/spatial-networks.tex
In \textit{spatially-embedded networks}, nodes are embedded in a metric space $\mathbf S$ with distance metric $d: \mathbf{S} \times \mathbf{S} \to \mathbb R^+$, and the presence of an edge between a pair of nodes $u$ and $v$ depends on the distance $d(\xu, \xv)$ between their embeddings $\xu$ and $\xv$, which for convenience we sometimes denote by $d(u, v)$. The probability of an edge between nodes $u$ and $v$ is determined by a connectivity probability function $\gamma : \posR \to [0,1]$, which commonly decays with distance. 

%% file: section/2-preliminaries/affiliation-based-networks.tex
An \textit{affiliation-based network} is a bipartite network in which nodes connect to affiliations. The bipartite network $H$ on $N$ nodes and $K$ affiliations is the tuple $(V, \mathcal A, \mathcal E)$ where $V$ is the node set $\{u \given u \in \{1, \dots N \}\}$, $\mathcal A$ is the set of affiliation nodes $\{a_i \given i \in \{1, \dots, K\}\}$, and edges in $\mathcal E \subseteq V \times \mathcal A$ connect individuals to affiliations.  The network $G = (V, E)$ is a projection of $H=(V, \mathcal A, \mathcal E)$ where $(u, v) \in E$ iff $\exists~ a_i$ such that $(u, a_i), (v, a_i) \in \mathcal E$, i.e. if $u$ and $v$ are co-affiliated with $a_i$ in $H$.

%% file: section/3-related-work/related-work.tex
In this section we describe the construction method and use of register-based social networks. We review existing random graph models and their limitations in the context of register-based social network data. In particular, we focus on models which are multiplex, have a spatial-embedding and/or have an affiliation-based connection mechanism. 

\subsection{Register-based social networks}
\label{subsec:register-based-social-networks}
\input{section/3-related-work/register-based-social-networks}

\subsection{Existing network models}
\label{subsec:existing-network-models}
\input{section/3-related-work/existing-models}

%% file: section/3-related-work/register-based-social-networks.tex
Register-based social networks have recently been constructed and studied in the Netherlands \cite{netherlands-anatomy}, Denmark \cite{danish-anatomy} and Sweden \cite{swedish-anatomy}. Data is compiled from various government-curated registers and commonly represents five types of relationships, namely \textit{family}, \textit{household}, \textit{next-door neighbours}, \textit{work} and \textit{school}, each represented as a layer in the network. Three of those five layers are unipartite projections of bipartite networks: in the \textit{work} and \textit{school} layers, individuals are connected if they are affiliated with the same employer or educational programme, and in the \textit{household} layer every individual living at the same address is connected. Sometimes node degrees are limited by only connecting to a subset of co-affiliated nodes determined by some other factor, commonly spatial proximity. In the \textit{next-door neighbour} layer individuals are connected to everyone in the ten geographically closest households to their address, which describes a spatially-embedded, almost bipartite structure, where individuals are connected to addresses but addresses can also connect to each other. The \textit{family} is the only layer in which connections are specifically pairwise, and individuals connect to their parents, siblings, partners, children and extended family.

Due to the data construction method~\cite{popnet-data}, many layers of a register-based network are at least approximately a unipartite projection of a bipartite network where each node connects to exactly one affiliation. The networks are multiplex as each layer is defined on the same set of nodes, and spatial proximity is accounted for explicitly in some layers (household, neighbourhood) and implicitly in others (work, school). These features introduce properties which are symptomatic of the construction method rather than the underlying social system, including high clustering within layers and a small range of node degrees. Thus these features should be accounted for in a relevant random network model.

The structural properties of register-based networks can be applied to the study of various social phenomena, including segregation~\cite{yuliiasocnets}, economic mobility, opinion formation~\cite{yuliia-scirep}, and epidemic spread~\cite{kieran-aer-insights,javier}.
Given its relevance for validating the MARS framework proposed in this paper, we turn our focus to one particular work, namely that of
\textcite{fragmentation}. It quantifies social cohesion, geographic mobility and the amount  of overlapping contacts across social contexts as properties of the ego
network of an individual $u$. Social cohesion is measured by network closure, which in particular we quantify as the local clustering coefficient
\begin{equation}
    c(u) = \frac{2 \tau(u)}{\deg_G(u)(\deg_G(u)-1)},
    \label{eq:local-clustering}
\end{equation}
where $\deg_G(u)$ is the degree of $u$ in $G$ and $\tau(u)$ is the number of triangles containing $u$. Geographic mobility is reflected by how spatially dispersed a node's connections are, which, in line with \textcite{fragmentation}, we quantify in the network as the average distance between alter pairs, given by
\begin{equation}
    d_{\text{alter}}(u) = \frac{\sum_{v,w \in N_G(u)}d(v,w)}{\deg_G(u)(\deg_G(u)-1)},
    \label{eq:alter-distance}
\end{equation}
where $N_G(u)$ is the set of alters of $u$. Finally, the amount of overlapping contacts across social contexts captures how common it is for individuals to be connected through more than one type of relationship (e.g. family members who are also coworkers). We will sometimes refer to this simply as the \textit{overlap in social contexts}. The multiplexity $m_{uv}$ of an edge $(u, v)$ is the number of layers in which the edge exists, i.e. $m_{uv} = \sum_{\ell = 1}^L|\{(u, v) \in E^\ell\}|$. Then the overlap in social contexts for a node $u$ can be measured by its share of multiplex edges
\begin{align}
    \mu_u = \frac{|\{(u, v): m_{uv} > 1\}|}{k_u}.
    \label{eq:multiplexity}
\end{align}
The global equivalents of these properties for a network $G$ are found by taking the average value over all ego networks. 
In their analysis of the Dutch register data, 
\textcite{fragmentation} show that an increase in mobility correlates to a decrease in overlap between social contexts and a decrease in social cohesion in the population-scale social network of the Netherlands. It is this particular empirical finding that we will revisit in our experiments using the MARS framework (Section \ref{sec:experiments}). 

%% file: section/3-related-work/existing-models.tex
Many elementary random network models, like Erdős–Rényi and Barabási–Albert graphs, do not capture the data peculiarities of register-based social networks discussed in Section \ref{subsec:register-based-social-networks}. Therefore, assumptions made based on such models will disproportionately account for these unusual intra-layer structures rather than the interesting social phenomena captured by the data. Thus, there is a need to move towards more complex, realistic network models which account for the multiplex, spatially-embedded and affiliation-based properties of the data.

Several network models which capture simple, space-dependent connection mechanisms have been well-studied. The Watts-Strogatz model approximates a spatial embedding by connecting nodes to their nearest neighbours in a ring topology, and generates ensembles by introducing random edge swaps. The model is designed to reproduce the ``small-world'' property of networks, but is not reflective of spatial dimensions in the real-world. Random Geometric Graphs (RGGs) are a class of network models where the existence of an edge between two nodes is a function of the distance between them in some embedding in a metric space $\mathbf S$, typically the unit square $[0,1]^2$. Classically, two nodes are adjacent in an RGG if they are within a radius $r$ of each other, corresponding to a connection probability step function where the probability of connection is 1 for distances less than or equal to $r$, and 0 otherwise. A more generalised version of the RGG is the soft RGG or Waxman model \cite{waxman}, in which nodes $u$ and $v$ are connected according to a probability
\begin{equation}
    p_{uv} = \beta e^{-\frac{d(u, v)}{\alpha \cdot r_0}},
    \label{eq:waxman-connection-function}
\end{equation}
which decays with distance, where $d(u, v)$ is the distance between $u$ and $v$ in the metric space $\mathbf S$, $r_0$ is the largest possible distance in $\mathbf S$, and $\alpha$ and $\beta$ are normalising parameters. Soft RGGs introduce a second level of randomness to classic RGGs: first in the embedding of the nodes in the metric space, and secondly in the assignment of edges. 

\textcite{multilayer-RGG} study an extension of the RGG to the multiplex context by generating $L$ RGGs independently on the same set of nodes with the same parameter $r$, and defining $G$ as the composition of these layers. However, this implies independence of the spatial dimension across layers, which is not reflective of the real-world system we want to capture (in the context of register-based social networks, an individual's embedding is their home address). Therefore, we need to move to a definition of a random graph where node positions are consistent and the assignment of edges introduces randomness across layers. 

Additionally, the $k$-connected $AB$ random geometric graph model is a bipartite extension of the RGG model. Type $A$ and type $B$ nodes are embedded in a metric space $S$, and each type $A$ node connects to the $k$ geographically closest type $B$ nodes \cite{AB-random-graph}. A single affiliation-based structure like a register-based network would therefore be represented by an $1$-connected $AB$ RGG. In the real-world social system, an individual's choice of affiliation depends on space but may not be strictly limited to the closest affiliation. Therefore, to extend such a model to this setting we need to introduce a probabilistic connection method which reflects this uncertainty in choice and also introduces randomness across layers. 

Due to the variation in definitions of register-based social networks across applications, a model for such data should be flexible.
\textcite{SERNs} introduce a framework for spatially-embedded random networks (SERNs) defined by a tuple $(\mathbf{S}, \gamma, X, N)$ where $N$ is the number of nodes, $\mathbf{S}$ is a metric embedding space, each node is embedded in $\mathbf{S}$ according to random variable $X$ and $\gamma: \mathbb R^+ \to [0,1]$ is a connectivity function which maps a distance $s$ between a pair of nodes in metric space $\mathbf{S}$ to a probability of them sharing an edge. \textcite{MLSERNs} extend this framework to multi-layer graphs (MLSERNs), such that nodes can be embedded in one or more of $L$ layers with a matrix of connection functions $\gamma_{\alpha\beta} : \mathbb R^+ \to [0,1]$ describing the probability of edges between nodes in layers $\alpha$ and $\beta$. To extend a framework like this to the context of register-based social networks, we need to consider multiplex networks with dependence between the spatial dimension in each layer, and an affiliation-based bipartite connection mechanism. To satisfy these conditions, we propose the MARS model.

%% file: section/4-proposed-approach/proposed-approach.tex
In this section we introduce MARS, a general network model which is multiplex, spatially-embedded and affiliation-based, as in register-based social networks. The governing principle behind the model is that nodes and affiliations are randomly embedded in a metric space, and each node chooses one affiliation in each layer based on a connectivity function which decays with distance. In this way, the spatial generative process of the network mirrors the mechanism by which an individual's choice of affiliation (e.g. workplace, educational institution) in the real world is influenced by their geographical proximity, and the multiplex, affiliation-based construction method replicates the way such networks are built from register data.

\subsection{Notation}
\input{section/4-proposed-approach/notation}

\subsection{Model definition}
\input{section/4-proposed-approach/model-definition}

%% file: section/4-proposed-approach/notation.tex
Following the notation introduced in Section \ref{sec:preliminaries}, we denote by $\{\Gl\}_{\ell=1}^L$ a set of graphs representing each layer of the network, where each $\Gl$ is the unipartite projection of an affiliation-based bipartite network $H^\ell$ on $N$ nodes and $K^\ell$ affiliations. When the layer $\ell$ is clear from context, we will sometimes denote $K^\ell$ simply by $K$ for ease of reading. The degree of each node in $H^\ell$ is exactly one, (each node chooses one affiliation), and we denote by $\al{}(u)$ the affiliation of $u$ in layer $\ell$. Then two nodes $u$ and $v$ are adjacent in $\Gl$ if $\al{}(u) = \al{}(v)$, and if a node $u$ chooses affiliation $\al{i}$ we denote this by $\al{}(u) = \al{i}$. We mostly consider the monoplex representation of the network $G$, where nodes $u$ and $v$ are adjacent if they are adjacent in at least one layer $\Gl$.

%% file: section/4-proposed-approach/model-definition.tex
We consider ensembles of MARS networks: sets of MARS networks defined on a given parameter set. This approach accounts for the natural variation in individual instances of a graph introduced by the randomness in the model.

\begin{definition*}
    A \textup{Multiplex Affiliation-based Random Spatially-embedded (MARS)} network ensemble is specified by a tuple $(N, L, \mathcal K, (\mathbf{S}, d), \omega, \mathbf X_V, \mathbf X_{\mathcal A})$, defined by:
    \begin{itemize}
        \item A number of nodes $N$.
        \item A number of layers $L$.
        \item A number of affiliations per layer $\{K^\ell\}_{\ell=1}^L$.
        \item A metric space $\mathbf{S}$ with a distance metric $d: \mathbf{S} \times \mathbf{S} \to \mathbb R^+$.
        \item A connectivity function $\omega: \mathbb R^+ \to \mathbb R^+$ which is a function of distance.
        \item Random variables $\mathbf X_V$ and $\mathbf X_\mathcal A$ with sample space $\mathbf S$, which represent the embedding method of individual nodes and affiliations respectively. 
    \end{itemize}
\end{definition*}

A realisation of an MARS network is generated by first sampling independently $N$ times from $\mathbf X_V$ to find $N$ embeddings $\{\mathbf x_1, \mathbf x_2, \dots, \mathbf x_N\}$ in $\mathbf S$ for each of the $N$ nodes in $V = \{1, \dots, N\}$. For each layer $\ell$, $\mathbf X_{\mathcal A}$ is similarly sampled independently $K=K^\ell$ times to find embeddings $\{\mathbf x_{\al{1}}, \mathbf x_{\al{2}}, \dots, \mathbf x_{\al{K}}\}$ for affiliations $\mathcal A^\ell = \{\al{1}, \al{2}, \dots, \al{K}\}$. For each layer $\ell$ with a set of affiliation positions $\xalset{j}$ and node positions $\{\xv\}_{v=1}^N$ the connection probability $\gamma^\ell : \underbrace{\posR \times \dots \times \posR}_{K^\ell +1} \to [0,1] $ is defined by 
\begin{equation}
    \gamma^\ell \left(u, \al{i} \right)
    =
    \prob{\al{}(u) = \al{i} \given \xu, \{\mathbf x_{\al{j}}\}_{j=1}^K} 
    := \frac{\omega \big(d(\xu, \xal{i})\big)}{\sum_{j =1}^{K} \omega \big(d(\xu, \xal{j})\big)}.
\label{eq:weighted-distance-function}
\end{equation}
This probability distribution is sampled once to find an affiliation for each node. A bipartite graph $\Hl=(V, \mathcal A^\ell, \mathcal E^l)$ is defined on the set of nodes $V$, affiliations $\mathcal A^\ell$ and edges between nodes and their sampled affiliations $\mathcal E^\ell$. Each individual node connects to exactly one affiliation in layer $\ell$ and so $|\mathcal E^\ell| = N$. The layer network $\Gl$ is the projection of $\Hl$ where nodes $u \in V$ are connected to all other nodes $v \in V$ with $\al{}(u) = \al{}(v)$. The resulting MARS network has a multiplex representation $\{\Gl\}_{\ell =1}^L$ and a monoplex representation $G$ where $u \sim v$ if there exists $\ell \in 1, \dots, L$ such that $u \sim_\ell v$, as defined in Section \ref{sec:multiplex-networks}.

Randomness is introduced to the network in two stages: first by the embeddings of nodes and per-layer affiliations, and secondly by the assignment of nodes to affiliations, which corresponds to edge assignment. The definition of an MARS network assumes that nodes embeddings are identically and independently distributed in the metric space, and that this also holds for affiliations.
We assume that $\omega$ is a decaying function, and indeed adopt this assumption in this study, but do not explicitly eliminate the case where $\omega$ may not be monotonically decreasing, in line with \textcite{SERNs}.

%% file: section/5-statistical-properties/statistical-properties.tex
In this section we investigate structural network statistics - including degree distribution, density and clustering - for the multiplex and monoplex representations of the MARS model, $\{\Gl\}_{\ell=1}^L$ and $G$. We consider special cases of the connection function $\omega$, and in particular examine the limiting cases (i) where the connectivity function is entirely space independent and (ii) where nodes only connect to their strictly closest affiliation. 

\subsection{Intra-layer properties}
\label{subsec:intra-layer-properties}
\input{section/5-statistical-properties/intra-layer}

\subsection{Monoplex properties}
\label{subsec:monoplex-properties}
\input{section/5-statistical-properties/monoplex}

\subsection{Limiting cases}
\label{subsec:limiting-cases}
\input{section/5-statistical-properties/special-cases}

%% file: section/5-statistical-properties/intra-layer.tex
The statistical properties of a single layer $\Gl$ of a MARS network $\{\Gl\}_{\ell=1}^L$ follow directly from the distribution of sizes of affiliations $\alset{i}$. We denote by $\Al{i}$ the \textit{size} of affiliation $\al{i}$, where the size of an affiliation is the number of nodes $u$ with $\al{}(u) = \al{i}$ or equivalently the order of the clique in $\Gl$ formed by co-affiliation with $\al{i}$. The size of an affiliation $\al{i}$ is dependent on its position in $\mathbf S$, its relative position to the other affiliations  $\{\al{j}\}_{j \ne i}$ in $\Gl$, the positions of the nodes $\{\mathbf x_1, \dots, \mathbf x_N\}$ and the connection function $\omega$. As all affiliations $\al{i}$ are independently and identically distributed according to $\XA$, all $\Al{i}$ can be described by a random variable $\mathbf A^\ell$.

\paragraph{Average degree, density and number of edges.} A node $u$ with affiliation $\al{}(u) = \al{i}$ has degree $\Al{i} - 1$ in layer $\ell$. Thus the (expected) \textit{average degree} $\kappa^\ell$ of $\Gl$ is
\begin{align}
    \kappa^\ell = \frac{1}{N} \sum_{i =1}^{K} \Al{i} (\Al{i} - 1), && \overline \kappa ^\ell := \expected{\kappa^\ell} = \frac{K}{N} \cdot \expected{(\mathbf A^\ell)_2},
\label{eq:exp-degree-second-moment}
\end{align}
where $\expected{(\mathbf A^\ell)_2}$ is the second factorial moment of $\mathbf A^\ell$. The (expected) \textit{number of edges} is given simply by
\begin{align}
    \varepsilon^\ell = \frac N 2 \kappa^\ell,  
    && \overline \varepsilon^\ell := \expected{\varepsilon^\ell} = \frac K 2 \cdot  \expected{(\mathbf A^\ell)_2},
\label{eq:exp-edge-count-second-moment}
\end{align}
and the (expected) \textit{density} by
\begin{align}
    \rho^\ell = \frac{2\cdot  \varepsilon^\ell}{N(N-1)}
    =
    \frac{\kappa^\ell}{N-1},
    &&
    \overline \rho^\ell := \expected{\rho^\ell} 
    =
    \frac{K}{N(N-1)} \cdot \expected{(\mathbf A^\ell)_2}.
\label{eq:exp-density-second-moment}
\end{align}
\paragraph{Triangles and clustering coefficients.} Within a layer triadic structure is trivial because $\Gl$ is a composition of cliques. Thus, the \textit{global clustering coefficient} and the \textit{local clustering coefficients} of each node are 1, and the (expected) \textit{number of triangles} is
\begin{align}
    \tau^\ell = \sum_{i=1}^{K} {\Al{i} \choose 3}, 
    && \overline \tau ^\ell := \expected{\tau^\ell} =
    \frac{K}{6} \cdot \expected{(\mathbf A^\ell)_3},
\label{eq:one-dimensional-triangles-by-layer}
\end{align}
where $\expected{(\mathbf A^\ell)_3}$ is the third factorial moment of $\mathbf A^\ell$. Similarly, as a composition of disconnected cliques, distance measures are not informative on $\Gl$. All pairs of nodes $u$ and $v$ with $\al{}(u) = \al{}(v)$ are a distance of 1 apart, and all pairs $u$ and $v$ with $\al{}(u) \ne \al{}(v)$ are in different components of the network implying $d(u, v) = \infty$.

\paragraph{Affiliation sizes.} Because each node is independently and identically distributed in $\mathbf S$ according to $\XV$ and each node makes its choice of affiliation independently, the number of nodes which choose a given affiliation is the result of $N$ Bernoulli trials with probability $\mathbf P^\ell$, where $\mathbf P^\ell$ is a random variable depending on $\XA$, $\XV$ and $\omega$ which describes how likely each affiliation is to be chosen by a random node. Therefore, the affiliation sizes are distributed according to a mixed binomial distribution $\mathbf A^\ell \sim \text{Bin}(N, \mathbf P^\ell)$, for which the $r$-th factorial moment is \cite{second-factorial-moments}
\begin{equation*}
    \expected{(\mathbf A^\ell)_r} =\frac{N!}{(N-r)!} \cdot \expected{(\mathbf P^\ell)^r},
\end{equation*}
where $\expected{(\mathbf P^\ell)^r}$ are the $r$-th raw moments of $\mathbf P^\ell$. Thus from equations (\ref{eq:exp-degree-second-moment} -- \ref{eq:one-dimensional-triangles-by-layer}) we have

\begin{align*}
    \overline \kappa ^\ell 
    =
    \frac{K}{N} \cdot \expected{(\mathbf A^\ell)_2}
    =
    K \cdot (N-1) \cdot \expected{(\mathbf P^\ell)^2},
    &&
    \overline \varepsilon^\ell 
    =
    \frac K 2 \cdot  \expected{(\mathbf A^\ell)_2}
    =
    K \cdot  {N \choose 2} \cdot \expected{(\mathbf P^\ell) ^2},
    \\
    \overline \rho^\ell 
    =
    \frac{K}{N(N-1)} \cdot \expected{(\mathbf A^\ell)_2}
    =
    K \cdot \expected{(\mathbf P^\ell)^2},
    &&
    \overline \tau ^\ell 
    =
    \frac{K}{6} \cdot \expected{(\mathbf A^\ell)_3}
    =
    K \cdot {N \choose 3} \cdot \expected{(\mathbf P^\ell)^3}.
\end{align*}
Each of these expressions is intuitive under the interpretation of $\expected{(\mathbf P^\ell)^r}$ as the probability that a set of $r$ nodes choose the same given affiliation, and thus $K \cdot \expected{(\mathbf P^\ell)^r}$ as the probability that they choose the same affiliation.

\paragraph{Degree distribution.} A node $u$ has degree $\deg_{\Gl}(u) = k$ in layer $\ell$ iff it is part of a clique of size $k + 1$, or equivalently if it chooses an affiliation $\al{i}$ of size $\Al{i} = k+1$. 
Because a uniformly sampled node is $k+1$ times more likely to belong to an affiliation of size $k+1$, the \textit{degree distribution} of a layer $\Gl$ is
\begin{equation}
    \prob{\deg_{\Gl}(u) = k} = \frac{k+1}{N} \cdot K\cdot \prob{\mathbf A^\ell = k + 1}.
\label{eq:intra-layer-degree-dist}
\end{equation}

%% file: section/5-statistical-properties/monoplex.tex
In the monoplex representation of $G$ a node $u$ is adjacent to a node $v$, denoted $u \sim v$, if $u$ and $v$ are adjacent in at least one layer $\Gl, ~\ell = 1, \dots, L$. Conversely, $u \not \sim v$ in $G$ if $u \not\sim_\ell v$ for all layers $\ell$. Thus
\begin{equation}
    \prob{u \sim v} = \prob{\bigcup_{\ell=1}^L u \sim_\ell v} = 1 - \prob{\bigcap_{\ell=1}^L u \not \sim_\ell v}.
    \label{eq:monoplex-density}
\end{equation}
In terms of affiliations, this means that $u$ is adjacent to $v$ if they share the same affiliation in at least one layer, i.e.
\begin{equation*}
    \prob{u \sim v} = \prob{\bigvee_{\ell=1}^L \al{}(u) = \al{}(v)} = 1 - \prob{\bigwedge_{\ell=1}^L \al{}(u) \ne \al{}(v)}.
\end{equation*}

\paragraph{Average degree, density and number of edges.} The statistical properties of the monoplex network $G$ depend not only on affiliation sizes but also on the number of overlapping nodes between affiliations across layers. Let $|\al{}(u)|$ denote the size of node $u$'s affiliation in layer $\ell$, and $|a^{\ell_1}(u) \cap a^{\ell_2}(u)|$ denote the number of overlapping nodes between node $u$'s affiliations in layers $\ell_1$ and $\ell_2$. Then $u$ has \textit{degree} $|\al{}(u)| - 1$ in $\Gl$, and by the inclusion-exclusion principle
\begin{align*}
    \deg_{G}(u) &=
    \sum_{\ell=1}^L \big(|\al{}(u)|-1\big) - \sum_{1\le\ell_1<\ell_2\le L} \big(|a^{\ell_1}(u) \cap a^{\ell_2}(u)|-1\big) + \dots + (-1)^{L+1}  \big(|a^1(u) \cap \dots \cap a^{L}(u)|-1\big),
\end{align*}
in the monoplex network $G$. Similarly, by the inclusion-exclusion principle the \textit{number of edges} $\varepsilon$ in $G$ is
\begin{align*}
    \varepsilon &= \sum_{\ell=1}^L \left(\sum_{i=1}^{K^\ell} {\Al{i} \choose 2}\right) - \sum_{1 \le \ell_1 < \ell_2 \le L}^{L} \left(\sum_{i_1=1}^{K^{\ell_1}} \sum_{i_2=1}^{K^{\ell_2}} {|a^{\ell_1}_{i_1} \cap a^{\ell_2}_{i_2}| \choose 2}\right) + \dots + (-1)^{L+1} \sum_{i_1 = 1}^{K^1} \dots \sum_{i_L = 1}^{K^L}{|a^1_{i_1} \cap \dots \cap a^L_{i_L}|\choose 2 },
\end{align*}
where the \textit{average degree} is $\kappa = 2\varepsilon / N$ and the \textit{density} is $\rho = 2\varepsilon / N(N-1)$.

\paragraph{Triangles and clustering coefficients.} In a multiplex network it is useful to distinguish between triangles with edges in one, two or three layers. We refer to these respectively as one-, two- and three-dimensional triangles \cite{multiplex-triadic-structures}.

\begin{figure}[!h]
    \centering
    \input{tikz-figures/multiplex-triangles}
    \caption{A (i) one- (ii) two- and (iii) three-dimensional triangle in a multiplex network.}
    \label{fig:placeholder}
\end{figure}
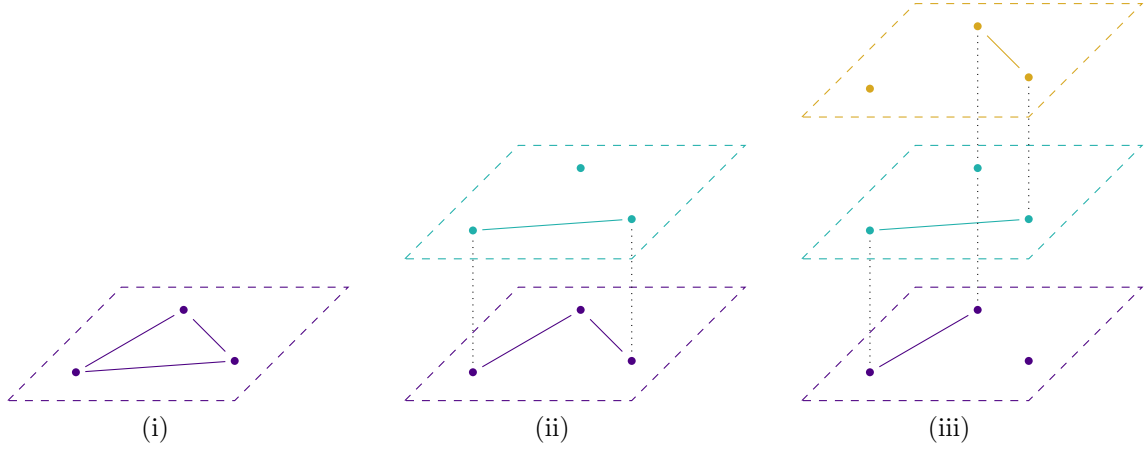

As discussed in Section \ref{subsec:intra-layer-properties}, triangle measures on a single layer $\Gl$ are not informative because of the composition of cliques structure. Therefore, we are more interested in triadic structures which form across multiple layers and which are not represented in any single layer. Under the MARS model, two-dimensional triangles necessarily have a one-dimensional equivalent, because the existence of edges $(u, v)$ and $(u, w)$ in layer $\ell$ implies the existence of edge $(v, w)$ by co-affiliation. Therefore, we only consider the number of one-dimensional triangles, which we denote by $\tau_{\text{1D}}$, and three-dimensional triangles which do not have a one-dimensional equivalent, which we denote by $\tau_{\text{3D}}$. The trio of nodes $u$, $v$ and $w$ can only be counted as a one- or three-dimensional triangle once, regardless of if it appears in multiple layers or combinations of layers, respectively.

Let $\tau^{\ell_1 \cap \ell_2}$ denote the \textit{number of one-dimensional triangles} which appear in both layers $\ell_1$ and $\ell_2$. Then 
\begin{equation}
    \tau_{\text{1D}} = \sum_{k = 1}^L (-1)^{k+1} \left ( 
    \sum_{1\le \ell_1 < \ell_2 < \dots < \ell_k \le L} \tau^{\ell_1 \cap \ell_2 \cap \dots \cap \ell_k}
    \right ).
\end{equation}
by the inclusion-exclusion principle.
The \textit{number of three-dimensional triangles} $\tau_{\text{3D}}$ is the number of node triplets where each edge appears in at least one layer, but no more than one edge appears in any layer.

We can also classify wedges as one-dimensional (both edges are in the same layer) or two-dimensional (each edge is in a different layer). The \textit{number of one-dimensional wedges} is three times the number of one-dimensional triangles
\begin{align}
    \nu_\text{1D} = 3 \cdot \tau_{\text{1D}}.
\end{align}
The \textit{number of two-dimensional wedges} $\nu_\text{2D}$ is the number of ordered node triplets $u$, $v$, $w$ where the two edges $(u, v)$ and $(v, w)$ appear in at least one layer, but never in the same layer. Each three-dimensional triangle contributes 3 two-dimensional wedges to the count $\nu_\text{1D}$, but, unlike in the one-dimensional case, there may also exist wedges which do not have triangle equivalents. The clustering coefficient of the network is given by
\begin{align*}
    C = 3 \cdot \frac{\tau_{\text{1D}} + \tau_{\text{3D}}}{\nu_{\text{1D}} + \nu_{\text{2D}}} = 3 \cdot \frac{\tau_{\text{1D}} + \tau_{\text{3D}}}{3 \cdot (\tau_{\text{1D}} + \tau_{\text{3D}} ) + (\nu_{\text{2D}} - 3 \cdot \tau_{\text{3D}})}
\end{align*}
and thus $C$ is controlled by the proportion of two-dimensional wedges which are not closed by any three-dimensional triangles. 

Let $\tau_{\text{1D}}(u)$ and $\tau_{\text{3D}}(u)$ denote the number of one- and three-dimensional triangles containing $u$, and denote by $\nu_{\text{1D}}(u)$ and $\nu_{\text{2D}}(u)$ the number of one- and two-dimensional wedges centred on $u$ (i.e. with edges $(u, v)$ and $(u, w)$ for some nodes $v$ and $w$). Then the number of one-dimensional wedges centred on $u$ is exactly the number of one-dimensional triangles containing $u$, i.e. $\nu_{\text{1D}}(u) = \tau_{\text{1D}}(u)$, and each three-dimensional triangle containing $u$ creates one two-dimensional wedge centred on $u$. Thus, as with the global clustering coefficient, the local clustering coefficient of $u$ is controlled by the proportion of two-dimensional wedges which are not closed by a three-dimensional triangle, 
\begin{align}
    c(u) = \frac{\tau_{\text{1D}}(u) + \tau_{\text{3D}}(u)}{\nu_{\text{1D}}(u) + \nu_{\text{2D}}(u)} = \frac{\tau_{\text{1D}}(u) + \tau_{\text{3D}}(u)}{(\tau_{\text{1D}}(u) + \tau_{\text{3D}}(u) ) + (\nu_{\text{2D}}(u) -  \tau_{\text{3D}}(u))}
\label{eq:monoplex-local-clustering}
\end{align}

%% file: tikz-figures/multiplex-triangles.tex
\begin{center}
\begin{tikzpicture}[scale=0.75, node/.style={draw, circle, inner sep=0pt, minimum size=.15cm, fill=black}]
    \draw[indigo, dashed] (0,0) -- (2,2) -- (6,2) -- (4,0) -- cycle;

    \fill[indigo] (1.2,0.5) circle (2pt) node (u1) {};
    \fill[indigo] (3.1,1.6) circle (2pt) node (v1) {};
    \fill[indigo] (4,0.7) circle (2pt) node (w1) {};

    \draw[indigo] (u1) -- (v1);
    \draw[indigo] (v1) -- (w1);
    \draw[indigo] (w1) -- (u1);

    \draw[indigo, dashed] (7,0) -- (9,2) -- (13,2) -- (11,0) -- cycle;
    \draw[lightseagreen, dashed] (7,2.5) -- (9,4.5) -- (13,4.5) -- (11,2.5) -- cycle;

    \fill[indigo] (8.2,0.5) circle (2pt) node (u1) {};
    \fill[indigo] (10.1,1.6) circle (2pt) node (v1) {};
    \fill[indigo] (11,0.7) circle (2pt) node (w1) {};

    \draw[indigo] (u1) -- (v1);
    \draw[indigo] (v1) -- (w1);

    \fill[lightseagreen] (8.2,3) circle (2pt) node (u2) {};
    \fill[lightseagreen] (10.1,4.1) circle (2pt) node (v2) {};
    \fill[lightseagreen] (11,3.2) circle (2pt) node (w2) {};

    \draw[lightseagreen] (w2) -- (u2);

    \draw[dotted] (u1) -- (u2);
    \draw[dotted] (w1) -- (w2);

    \draw[indigo, dashed] (14,0) -- (16,2) -- (20,2) -- (18,0) -- cycle;
    \draw[gold, dashed] (14,5) -- (16,7) -- (20,7) -- (18,5) -- cycle;
    \draw[lightseagreen, dashed] (14,2.5) -- (16,4.5) -- (20,4.5) -- (18,2.5) -- cycle;

    \fill[indigo] (15.2,0.5) circle (2pt) node (u1) {};
    \fill[indigo] (17.1,1.6) circle (2pt) node (v1) {};
    \fill[indigo] (18,0.7) circle (2pt) node (w1) {};

    \draw[indigo] (u1) -- (v1);

    \fill[lightseagreen] (15.2,3) circle (2pt) node (u2) {};
    \fill[lightseagreen] (17.1,4.1) circle (2pt) node (v2) {};
    \fill[lightseagreen] (18,3.2) circle (2pt) node (w2) {};

    \draw[lightseagreen] (w2) -- (u2);

    \fill[gold] (15.2,5.5) circle (2pt) node (u3) {};
    \fill[gold] (17.1,6.6) circle (2pt) node (v3) {};
    \fill[gold] (18,5.7) circle (2pt) node (w3) {};

    \draw[gold] (v3) -- (w3);

    \draw[dotted] (u1) -- (u2);
    \draw[dotted] (v1) -- (v2);
    \draw[dotted] (v3) -- (v2);
    \draw[dotted] (w3) -- (w2);

    \node at (2.6, -0.5) {(i)};

    \node at (9.6, -0.5) {(ii)};

    \node at (16.6, -0.5) {(iii)};
\end{tikzpicture}
\end{center}

%% file: section/5-statistical-properties/special-cases.tex
In the following sections we discuss two special cases of MARS: firstly where the connectivity function $\omega$ is entirely space-independent, and secondly where nodes only connect to their strictly closest affiliation. In the latter section we introduce the configuration used in our experiments in Section~\ref{sec:experiments}. 

\subsubsection{Space-independent case}
\label{subsubsec:space-independent-case}
In the space-independent case each node chooses their affiliation in each layer uniformly at random. Therefore, the probability of a node $u$ choosing any affiliation $\al{i}$ in layer $\ell$ is $\frac 1 {K^\ell}$ and $\mathbf A^\ell$ is described by a binomial distribution $\mathbf A^\ell \sim \text{Bin}\left(N, \frac{1}{K^\ell}\right)$ with second factorial moment $\expected{(\mathbf A^\ell)_2} = N(N-1)/({K^\ell})^2$. From Equations (\ref{eq:exp-degree-second-moment} - \ref{eq:exp-density-second-moment}) this gives
\begin{align*}
    \overline \kappa ^\ell =  \frac{K^\ell}{N} \cdot \expected{(\mathbf A^\ell)_2}
    =
    \frac{1}{K^\ell} \cdot (N-1),~~~~
    &&
    \overline \varepsilon^\ell =  
    \frac {K^\ell} 2 \cdot  \expected{(\mathbf A^\ell)_2}
    =
    \frac{1}{K^\ell} \cdot {N \choose 2},~~~~
    &&
    \overline \rho^\ell =
    \frac{K^\ell}{N(N-1)} \cdot \expected{(\mathbf A^\ell)_2} = \frac{1}{K^\ell}.
\end{align*}

In each of these first-order properties a single layer $\Gl$ of a space-independent MARS network is indistinguishable from an Erdős-Rényi $G(N, \frac 1 {K^\ell})$ random graph. Furthermore, each of these properties is minimised under the MARS model by the fact that $\sum_i ({\Al{i}})^2$ is minimised under the constraint $\sum_i \Al{i} = N$ when $\forall i \Al{i} = \frac{N}{K^\ell}$.

The third factorial moment of a binomial distribution $\mathbf A^\ell \sim \text{Bin}(N, \frac{1}{K^\ell})$ is $\expected{(\mathbf A^\ell)_3} = N(N-1)(N-2)/(K^\ell)^3$. Thus by (\ref{eq:one-dimensional-triangles-by-layer}) the number of triangles in a layer $\Gl$ of a space-independent MARS network is,
\begin{align}
    \overline \tau ^\ell = 
    \frac{K^\ell}{6} \cdot \expected{(\mathbf A^\ell)_3} 
    = 
    {N \choose 3}\cdot \frac 1 {(K^\ell)^2},
\label{eq:space-independent-layer-triangles}
\end{align}
which is a factor of $K^\ell$ larger than an Erdős-Rényi $G(N, \frac 1 {K^\ell})$ random graph. In this way, a single layer of a space independent MARS network is equivalent to an Erdős-Rényi graph with complete triadic closure.

Space independence means that each layer of the network can be modelled independently. A pair of nodes choose the same affiliation in each layer $\ell$ independently with probability $1/K^\ell$. Therefore (\ref{eq:monoplex-density}) the expected density of $G$ is
\begin{equation}
    \overline \rho = 
    1 - \prod_{\ell=1}^L \left( 1- \frac{1}{K^\ell}\right),
\label{eq:uniform-case-density}
\end{equation}
and values for $\overline \kappa$ and $\overline \varepsilon$ follow directly from $\overline \rho$. In terms of first-order properties, a MARS network $G$ is once again equivalent to an Erdős-Rényi $G(N, \overline \rho)$ random graph.

By a similar argument, a trio of nodes all choose the same affiliation in layer $\ell$ with probability $1/(K^\ell)^2$ and the expected number of unique one-dimensional triangles in $G$ is
\begin{equation}
    \overline \tau_{\text{1D}} = {N\choose 3} \left[1 - \prod_{\ell=1}^L\left(1 - \frac{1}{(K^\ell)^2} \right) \right].
\label{eq:uniform-case-1D-triangles}
\end{equation}

As discussed in Section \ref{subsec:monoplex-properties}, all two-dimensional triangles are included in (\ref{eq:uniform-case-1D-triangles}) and we only need to count the remaining three-dimensional triangles which do not have one-dimensional equivalents, given by (Appendix \ref{app:triangle-counting}) 
\begin{align}
    \overline \tau_{\text{3D}} = {N\choose 3} \left[ \prod_{\ell = 1}^L \left(1 - \frac{1}{(K^\ell)^2} \right)
    -
    3\prod_{\ell=1}^L \left(1 - \frac{1}{{K^\ell}}\right) 
    +
    3 \prod_{\ell=1}^L \left(\frac{K^\ell - 1}{K^\ell}\right)^2 
    -
    \prod_{\ell=1}^L \frac{(K^\ell -1)(K^\ell - 2)}{(K^\ell)^2} \right].
\label{eq:uniform-case-3D-triangles}
\end{align}

The number of one-dimensional wedges $\overline \nu_{\text{1D}}$ is simply given by
\begin{equation}
    \overline \nu_{\text{1D}} = 3 \cdot {N\choose 3} \left[1 - \prod_{\ell=1}^L\left(1 - \frac{1}{(K^\ell)^2} \right) \right].
\label{eq:uniform-case-1D-wedges}
\end{equation}
The number of two-dimensional wedges which do not have a one-dimensional equivalent is given by (Appendix \ref{app:triangle-counting}) 
\begin{align}
    \overline \nu_{\text{2D}} = 3 \cdot {N \choose 3} \left[
    \prod_{\ell = 1}^L \left(1 - \frac{1}{(K^\ell)^2} \right) 
    -
    2 \prod_{\ell = 1}^L \left(1 - \frac{1}{K^\ell} \right) 
    +
    \prod_{\ell = 1}^L \left (\frac{K^\ell - 1}{K^\ell} \right )^2 \right].
\label{eq:uniform-case-2D-wedges}
\end{align}
Thus, the global clustering coefficient of $G$ is
\begin{align}
    C(G) = 3 \cdot \frac{\overline \tau_{\text{1D}} + \overline \tau_{\text{3D}}}{\overline \nu_{\text{1D}} + \overline \nu_{\text{2D}}} 
    =
    \frac{ \left(1 
    -
    3\prod_{\ell=1}^L \left(1 - \frac{1}{{K^\ell}}\right) 
    +
    3 \prod_{\ell=1}^L \left(\frac{K^\ell - 1}{K^\ell}\right)^2 
    -
    \prod_{\ell=1}^L \left(\frac{(K^\ell - 1)(K^\ell - 2)}{(K^\ell)^2}\right) \right) }
     {\left(1 
    -
    2 \prod_{\ell = 1}^L \left(1 - \frac{1}{K^\ell} \right) 
    +
    \prod_{\ell = 1}^L \left (\frac{K^\ell - 1}{K^\ell} \right )^2 \right)}.
\label{eq:uniform-case-clustering}
\end{align}
By an identical argument, substituting $N \choose 3$ for $N-1 \choose 2$ in Equations (\ref{eq:uniform-case-1D-triangles} -- \ref{eq:uniform-case-2D-wedges}), we can show that the average local clustering coefficient is also given by (\ref{eq:uniform-case-clustering}).

In a single layer $\Gl$ the degree distribution (\ref{eq:intra-layer-degree-dist}) is given by 
\begin{equation*}
    \prob{\deg_{\Gl}(u) = k} = \frac{k+1}{N} \cdot K^\ell \cdot {N \choose k + 1} \left( \frac 1 {K^\ell} \right)^{k+1} \left(1- \frac 1 {K^\ell} \right)^{N-k-1}.
\end{equation*}
This is equivalent to, 
\begin{equation}
    \prob{\deg_{\Gl}(u) = k} = {N - 1 \choose k} \left( \frac 1 {K^\ell} \right)^{k} \left(1- \frac 1 {K^\ell} \right)^{N-k-1},
\label{eq:uniform-case-per-layer-degree-distribution}
\end{equation}
which is the degree distribution of an Erdős-Rényi $G(N, \frac 1 {K^\ell})$ random graph. Similarly, the degree distribution of the monoplex $G$ is given by, 
\begin{equation}
    \prob{\deg_{G}(u) = k} = {N - 1 \choose k} {\overline \rho}^{k} \left(1- \overline \rho\right)^{N-k-1},
\label{eq:uniform-case-degree-distribution}
\end{equation}
where $\overline \rho$ is defined in Equation (\ref{eq:uniform-case-density}).

\begin{figure}[!h]
    \centering
    \begin{subfigure}[t]{0.335\linewidth}
        \centering
        \includegraphics[width=\linewidth]{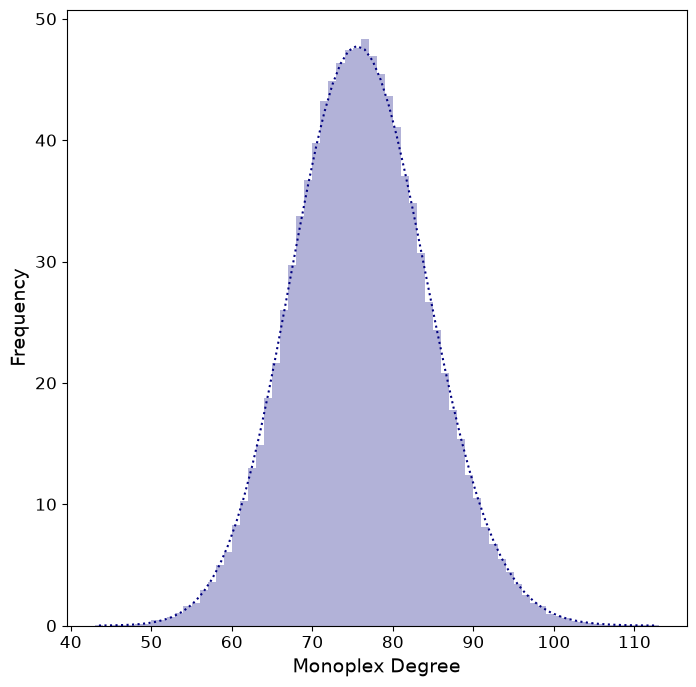}
        \caption{}
        \label{fig:space-independent-analytical-propertiesA}
    \end{subfigure}
    \hspace{1.2em}
    \begin{subfigure}[t]{0.335\linewidth}
        \centering
        \includegraphics[width=\linewidth]{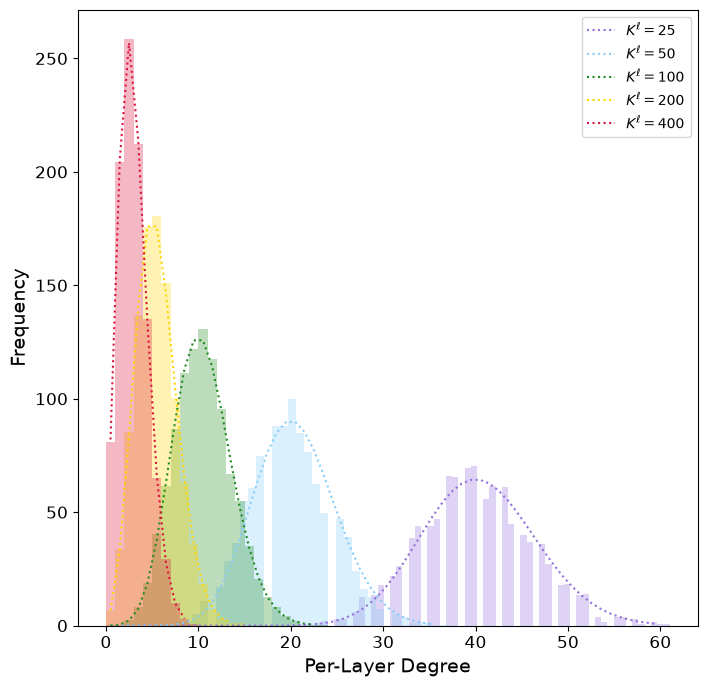}
        \caption{}
        \label{fig:space-independent-analytical-propertiesB}
    \end{subfigure}
    \begin{subfigure}[t]{0.22\linewidth}
        \centering
        \includegraphics[width=\linewidth]{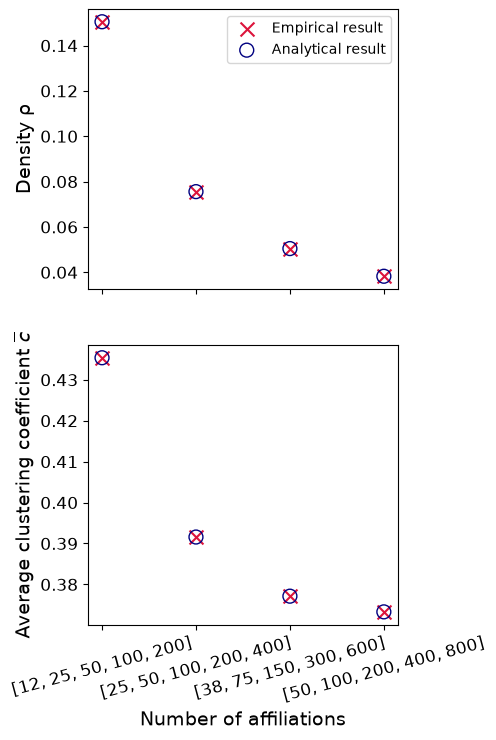}
        \caption{}
        \label{fig:space-independent-analytical-propertiesCD}
    \end{subfigure}
    
    \caption{A comparison of the analytically derived properties of a space-independent MARS network against experimental results averaged over 100 runs. (a) The degree distribution of the monoplex network $G$ with 1000 nodes and 5 layers with 25, 50, 100, 200 and 400 affiliations is a binomial distribution with probability $\rho 
\approx 0.07558$ given by (\ref{eq:uniform-case-density}). (b) The per-layer degree distribution of layers $\Gl$ on 1000 nodes with $K^\ell = $ 25, 50, 100, 200 and 400 affiliations are binomial distributions with probability $1/K^\ell$. (c) The density $\rho$ and average local clustering coefficient $\overline c$ of MARS networks on 1000 nodes for varying numbers of affiliations. The analytical results are given by (\ref{eq:uniform-case-density}), (\ref{eq:uniform-case-1D-triangles}) and (\ref{eq:uniform-case-3D-triangles}).}
    \label{fig:space-independent-analytical-properties}
\end{figure}

\subsubsection{Hard geometric case}
\label{subsubsec:hard-geometric-case}
When $\omega$ is such that in each layer all nodes necessarily choose their strictly closest affiliation, each layer is a unipartite projection of a $1$-connected $AB$ random geometric graph \cite{AB-random-graph}. The affiliation embeddings $\xalset{i}$ are the nuclei of a Voronoi decomposition of the metric space $\mathbf S$, and each node connects to all other nodes in the same Voronoi cell. The size of each affiliation $\al{i}$ is determined by the volume of the Voronoi cell with nucleus $\xal{i}$, which we denote by $M^\ell_i$, and the density of $\XV$ in that cell. If we describe the volume distribution of Voronoi cells by a random variable $\mathbf M^\ell$ which depends on $\mathbf S$ and $\XA$, then $\mathbf P^\ell$ is a random variable dependent on $\mathbf M^\ell$ and $\XV$. 

Because the properties of a MARS network under the hard geometric condition are strongly dependent on $\mathbf S$, $\XV$ and $\XA$, in this section we consider specific values of these parameters. In particular, we assume $\mathbf S$ to be the unit square $[0,1]^2$ and affiliation embeddings in layer $\Gl$ to be modelled by a Poisson Point Process (PPP) with intensity $K^\ell$ per unit area. Accordingly, nodes and affiliations have two-dimensional embeddings
\begin{align*}
    \{\mathbf x_u\}_{u=1}^N = \{(x_1, y_1), (x_2, y_2), \dots, (x_N, y_N)\},~~
    &&
    \{\mathbf x_{\al{i}}\}_{i=1}^{K} =\{(x_{\al{1}}, y_{\al{1}}), (x_{\al{2}}, y_{\al{2}}), \dots, (x_{\al{K}}, y_{\al{K}})\}
\end{align*}
 for $\ell = 1, \dots, L$, respectively, and the metric $d$ is Euclidean distance.  In this way, we can derive a mathematical approximation for the simple case where $K^\ell$ affiliations per layer $\Gl$ are embedded uniformly in the unit square $[0,1]^2$, which we explore in Section \ref{sec:experiments}. 

\paragraph{Uniform Node Embeddings}

First we consider the case where both nodes and affiliations are distributed in $[0,1]^2$ according to independent PPPs with intensities $N$ and $\{K^\ell\}_{\ell=1}^L$ for each layer, respectively. This is an approximation of the case where nodes are also embedded uniformly in $\mathbf S$. 

The probability that a node falls in a given Voronoi cell with nucleus $\xal{i}$ is exactly its area $M^\ell_i$. Therefore $\mathbf P^\ell = \mathbf B^\ell$ and it is known that the second moment of the cell area of a Poisson-Voronoi decomposition with intensity $K$ is approximately $1.280 / K^2$ \cite{voronoi-polygon, poisson-voronoi-distribution}. Therefore by Equations (\ref{eq:exp-degree-second-moment} - \ref{eq:exp-density-second-moment})
\begin{align*}
    \overline \kappa ^\ell =  \frac{K^\ell}{N} \cdot \expected{(\mathbf A^\ell)_2}
    \approx
    \frac{1.28}{K^\ell} \cdot (N-1),~~
    &&
    \overline \varepsilon^\ell =  
    \frac {K^\ell} 2 \cdot  \expected{(\mathbf A^\ell)_2}
    \approx
    \frac{1.28}{K^\ell} \cdot {N \choose 2},~~
    &&
    \overline \rho^\ell =
    \frac{K^\ell}{N(N-1)} \cdot \expected{(\mathbf A^\ell)_2} \approx \frac{1.28}{K^\ell}.
\end{align*}
The third moment of the cell area of a Poisson-Voronoi decomposition with intensity $K$ in the plane $\sqR$  has been approximated as $2.020 / K^3$ \cite{poisson-voronoi-distribution}. Therefore
\begin{align}
    \overline \tau ^\ell 
    =
    \frac{K}{6} \cdot \expected{(\mathbf A^\ell)_3}
    \approx
    {N \choose 3} \cdot \frac{2.02}{(K^\ell)^2}.
    \label{eq:hard-geometric-layer-triangles}
\end{align}
We expect that with boundary conditions these values will increase due to the added variance in Voronoi cell areas introduced by boundary cells with no competing nuclei outside the unit square.

In the hard geometric case, nodes' choices of affiliations are conditionally dependent on their embeddings, and 
\begin{align*}
\prob{u \sim v ~;~ \xu, \xv, \xalset{j}}
=
\sum_{i = 1}^{K^\ell} \prob{\al{}(u) = \al{i} \cap \al{}(v) = \al{i} ~;~ \xu, \xv, \xalset{j}}.
\end{align*}
Given that affiliation embeddings are independent and identically distributed, the expected probability that two nodes at positions $\xu$ and $\xv$ connect is
\begin{align*}
    \prob{u \sim v ~;~ \xu, \xv}
    =
    K^\ell \cdot \int_{\mathbf S}\prob{\al{}(v) = \al{i} \given \al{}(u) = \al{i}  ~;~ \xu, \xv}\cdot \prob{\al{}(u) = \al{i} ~;~ \xu} \dx{\xal{i}} 
\end{align*}
A node $u$ chooses an affiliation $\al{i}$ if there exists no other affiliation $\al{j}$ such that $d(\xu, \xal{j}) < d(\xu, \xal{i})$, and two nodes $u$ and $v$ connect in layer $\ell$ if there exists an affiliation $\al{i}$ which satisfies this condition for both nodes. The probability that this happens is equivalent to the void probability of the PPP with intensity $K^\ell -1$ over the union of the balls $B_{d(u, \al{i})}(\xu)$ and $B_{d(v, \al{i})}(\xv)$ passing through $\xal{i}$ centred at $\xu$ and $\xv$ respectively, given by
\begin{align}
     \prob{u \inl v \given \xu, \xv} = K^\ell \cdot \int_{\mathbf S} \exp \big(- (K^\ell -1) \cdot |B_{d(u, \al{i})}(\xu) \cup B_{d(v, \al{i})}(\xv) |\big) \ \dx{\xal{i}}.
\label{eq:voronoi-connect-prob-given-embeddings}
\end{align}
Note that this integral is complicated because of the various ways two circles can overlap in the unit square.
The probability that two nodes distance $\delta$ apart connect in layer $\Gl$ is given by integrating (\ref{eq:voronoi-connect-prob-given-embeddings}) over all such pairs of embeddings
\begin{equation}
    \prob{u \inl v \given d(u, v) = \delta} = \iint_{\substack{\mathbf S^2, d(\xu, \xv) = \delta}} \prob{u \inl v \given \xu, \xv} \dx{\xu} \dx{\xv},
\label{eq:distance-connection-prob-voronoi}
\end{equation}
and the expected density of the monoplex network $G$ is given by the integral
\begin{align*}
    \overline \rho = \iint_{\mathbf S^2} \left[1 - \prod_{\ell = 1}^L \big ( 1- \prob{u \inl v \given \xu, \xv} \big) \right] \dx{\xu} \dx{\xv}.
\end{align*}

\begin{figure}[!h]
    \centering
    \includegraphics[width=0.5\linewidth]{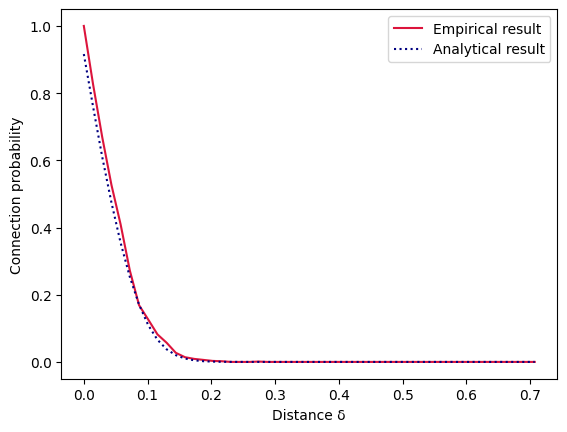} 
    \caption{The relationship between pairwise distance $\delta$ between nodes and their probability of connecting. The red line shows the average of 1000 runs on a network with 1000 nodes and 5 layers with 25, 50, 100, 200 and 400 affiliations. The dotted line shows the analytical value given by (\ref{eq:distance-connection-prob-voronoi}).}
    \label{fig:placeholder}
\end{figure}

\begin{figure}[!h]
    \centering
    \begin{subfigure}[t]{0.335\linewidth}
        \centering
        \includegraphics[width=\linewidth]{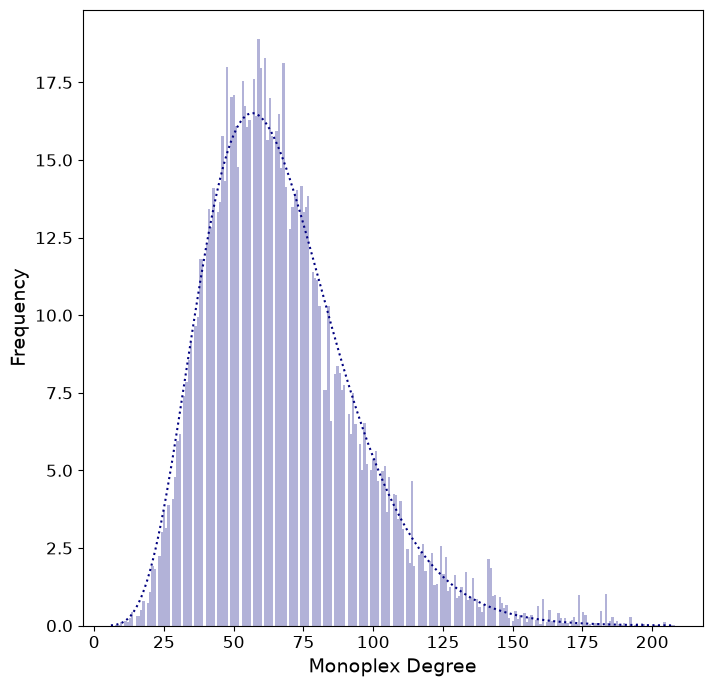}
        \caption{}
        \label{fig:hard-geometric-analytical-propertiesA}
    \end{subfigure}
    \hspace{1.2em}
    \begin{subfigure}[t]{0.335\linewidth}
        \centering
        \includegraphics[width=\linewidth]{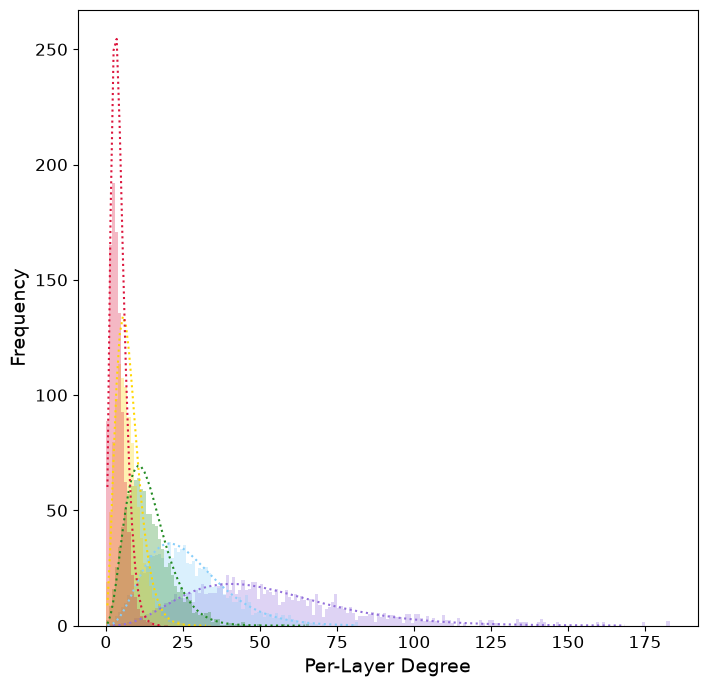}
        \caption{}
        \label{fig:hard-geometric-analytical-propertiesB}
    \end{subfigure}
    \begin{subfigure}[t]{0.22\linewidth}
        \centering
        \includegraphics[width=\linewidth]{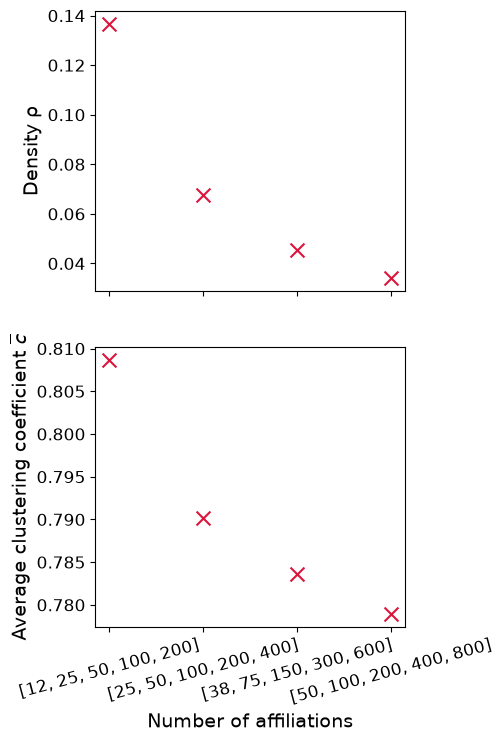}
        \caption{}
        \label{fig:hard-geometric-analytical-propertiesCD}
    \end{subfigure}
    
    \caption{A comparison of the analytically derived properties of a hard geometric MARS network against experimental results averaged over 100 runs. (a) The degree distribution of the monoplex network $G$ with 1000 nodes and 5 layers with 25, 50, 100, 200 and 400 affiliations. The distribution is well approximated by the distribution $\text{Gamma}(6.67, 0.1)$. (b) The per-layer degree distribution of layers $\Gl$ on 1000 nodes with $K^\ell = $ 25, 50, 100, 200 and 400 affiliations are approximately gamma distributions described by (\ref{eq:hard-geometric-layer-degree-dist}) (c) The density $\rho$ and average local clustering coefficient $\overline c$ of MARS networks on 1000 nodes for varying numbers of affiliations.}
    \label{fig:hard-geometric-analytical-properties}
\end{figure}

By a similar argument to above
\begin{align*}
    \prob{u \inl v \inl w ~;~ \xu, \xv, \mathbf x_w} = K^\ell \int_{\mathbf S} \exp \big(- (K^\ell -1) \cdot |B_{d(u, \al{i})}(\xu) \cap B_{d(v, \al{i})}(\xv) \cap B_{d(w, \al{i})}(\mathbf x_w)|\big) \dx{\xal{i}},
\end{align*}
and thus the expected number of one-dimensional triangles is
\begin{align*}
    \overline \tau_{\text{1D}}  = \iiint_{\mathbf S^3} \left [ 1 -  \prod_{\ell = 1}^L \big ( 1 - \prob{u \inl v \inl w ~;~ \xu, \xv, \mathbf x_w}\big) \right] \dx{\xu} \dx{\xv} \dx{\mathbf x_w}.
\end{align*}
The expected number of one-dimensional wedges is $\overline \nu_{\text{1D}} = 3 \cdot \overline \tau_{\text{1D}}$. The analytical derivation of the expected number of three-dimension triangles $\overline \tau_{\text{3D}}$ and two-dimensional wedges $\overline \nu_{\text{2D}}$ is complex and beyond the scope of this paper, but we present empirical results for the clustering coefficient in Figure \ref{fig:hard-geometric-analytical-propertiesCD}.

The affiliation size distribution in layer $\Gl$ is approximated by \cite{AB-random-graph}
\begin{equation*}
    \prob{\mathbf A^\ell = k} \approx \frac{\Gamma(k + \alpha)}{\Gamma(k + 1)\Gamma(\alpha)} \cdot \frac{\alpha^\alpha (N/K)^k}{(N/K + \alpha)^{k + \alpha}},
\end{equation*} 
where $\alpha = 3.57$ \cite{poisson-voronoi-distribution, voronoi-cell-size}. Thus, by Equation (\ref{eq:intra-layer-degree-dist}), the degree distribution of $\Gl$ is 
\begin{equation}
    \prob{\deg_{\Gl}(u) = k} \approx \frac{k+1}{N} \cdot K^\ell \cdot \frac{\Gamma(k + 1 + \alpha)}{\Gamma(k + 2)\Gamma(\alpha)} \cdot 
    \frac{\alpha^\alpha (N/K)^{k+1}}{(N/K + \alpha)^{k + 1 + \alpha}}.
\label{eq:hard-geometric-layer-degree-dist}
\end{equation} 

The analytical derivation of the monoplex degree distribution of the network is not straightforward, but Figure \ref{fig:hard-geometric-analytical-propertiesA} shows that it is well approximated by a gamma distribution with $\alpha = 6.67$ and $\beta = 0.1$. The gamma distribution is fitted to the experimental degree distribution by minimising the mean squared error. 

Figures \ref{fig:space-independent-analytical-propertiesCD} and \ref{fig:hard-geometric-analytical-propertiesCD} show that the density of the MARS network in the hard geometric case with uniform node embeddings is lower than in the space-independent case. Although intra-layer densities are higher in the hard geometric case, nodes are likely to connect to the same proximate nodes in multiple layers. In the space independent case, the likelihood of the same edge forming in more than one layer in very small and so the number of unique edges is higher, leading to a higher density. 

We also see that the average local clustering coefficient is much higher for uniform node embeddings in the hard geometric case than in the space independent case. By Equation (\ref{eq:monoplex-local-clustering}) we know that the local clustering coefficient of a node is mostly controlled by the proportion of two-dimensional wedges which are not closed by three-dimensional triangles. In the space independent case edges are independent across layers, and so the probability that an open two-dimensional wedge is closed is very small. In the hard geometric case, triplets of nodes which are spatially close are likely to be connected, and so we expect to see high rates of inter-layer triadic closure. 

\paragraph{Gaussian Node Embeddings} We also consider the case where nodes are distributed normally around a mean $(0.5, 0.5)$ with standard deviation $\sigma$. This approximates the case where nodes are embedded according to a two-dimensional truncated normal distribution with independence between the dimensions. 

\begin{figure}[!h]
    \centering
    \begin{subfigure}[t]{0.335\linewidth}
        \centering
        \includegraphics[width=\linewidth]{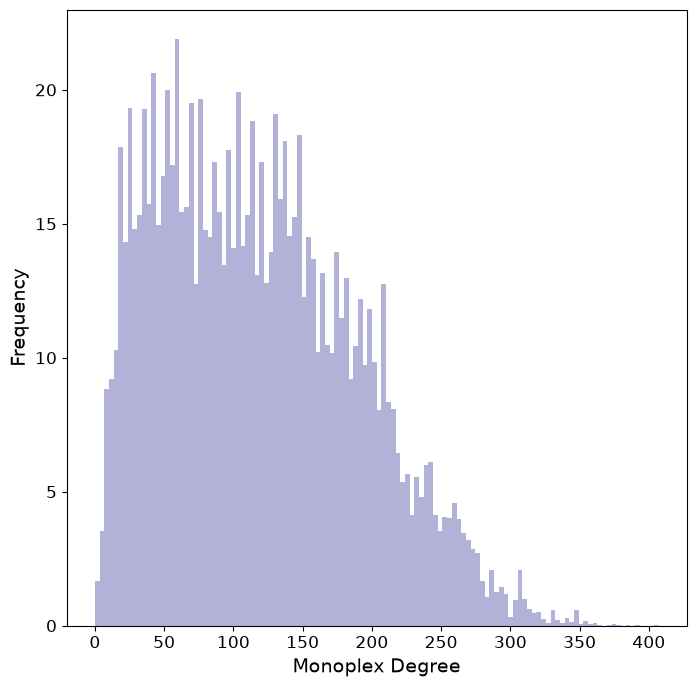}
        \caption{}
        \label{fig:normal-embeddings-analytical-propertiesA}
    \end{subfigure}
    \hspace{1.2em}
    \begin{subfigure}[t]{0.335\linewidth}
        \centering
        \includegraphics[width=\linewidth]{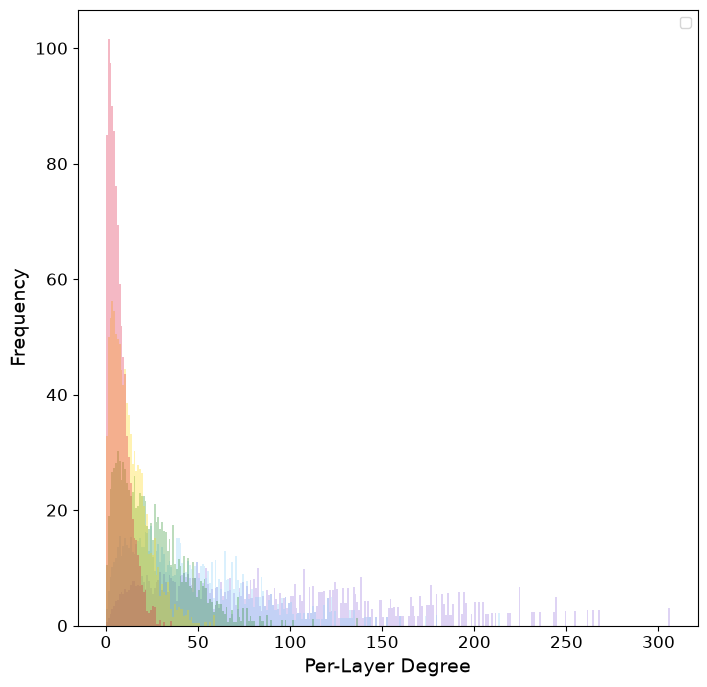}
        \caption{}
        \label{fig:normal-embeddings-analytical-propertiesB}
    \end{subfigure}
    \begin{subfigure}[t]{0.22\linewidth}
        \centering
        \includegraphics[width=\linewidth]{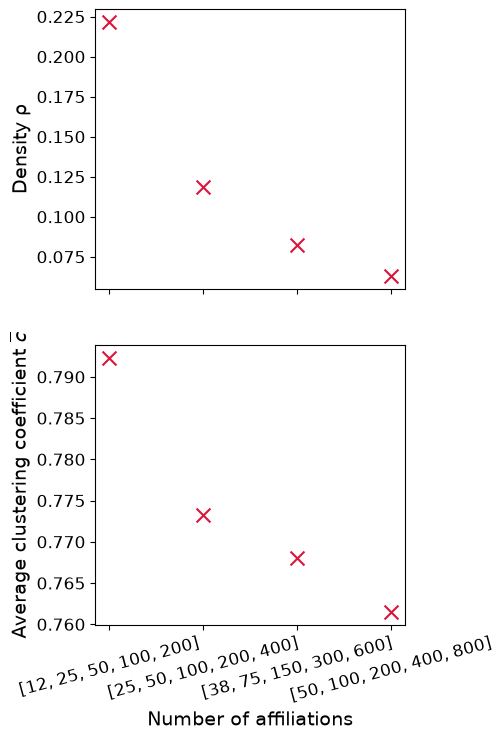}
        \caption{}
        \label{fig:normal-embeddings-analytical-propertiesCD}
    \end{subfigure}
    
    \caption{The properties of a hard geometric MARS network with Gaussian node embeddings centred around a mean $(0.5, 0.5)$ with standard deviation $\sigma = 0.2$, averaged over 100 runs. (a) The degree distribution of the monoplex network $G$ with 1000 nodes and 5 layers with 25, 50, 100, 200 and 400 affiliations. (b) The per-layer degree distribution of layers $\Gl$ on 1000 nodes with $K^\ell = $ 25, 50, 100, 200 and 400 affiliations. (c) The density $\rho$ and average local clustering coefficient $\overline c$ of MARS networks on 1000 nodes for varying numbers of affiliations.}
    \label{fig:normal-embeddings-analytical-properties}
\end{figure}

We expect the intra-layer density $\rho^\ell$, number of edges $\varepsilon^\ell$ and average degree $\kappa^\ell$ to be higher than in the case of uniform node embeddings because of the increased variation in affiliation size caused by the increased variation in node density. The analytical derivations of these values are complex and outside the scope of this paper. The density of the monoplex representation $G$ is given by
\begin{align*}
    \overline \rho = \iint_{\mathbf S^2} \left[1 - \prod_{\ell = 1}^L \big ( 1- \prob{u \inl v \given \xu, \xv} \big) \right] \cdot \pdf{\XV}(\xu) \cdot \pdf{\XV}(\xv) \dx{\xu} \dx{\xv},
\end{align*}
where $\pdf{\XV}(\xu) = \frac{1}{2\pi\sigma^2} \exp\Big({{-\frac{1}{2\sigma^2}}\left ((x_u - 0.5)^2 + (y_u - 0.5)^2 \right)}\Big)$ is the probability density of a two-dimensional normal distribution for $\xu = (x_u, y_u)$ and values for $\overline \kappa$ and $\overline \varepsilon$ follow directly from $\overline \rho$. 
Due to the complexity of the MARS model in the Gaussian node embedding case, we explore its properties through simulations and show the results in Figure \ref{fig:normal-embeddings-analytical-properties}.

Figure \ref{fig:normal-embeddings-analytical-propertiesA} shows that the degree distribution is a skewed distribution like the gamma distribution for uniform node embeddings (Figure \ref{fig:hard-geometric-analytical-propertiesA}) but with more nodes with higher degrees. We expect nodes with embeddings near the centre of the unit square to have significantly higher degrees in the case of normally embedded nodes, because affiliations near the centre will be large. The intra-layer degree distributions show a similar pattern (Figure \ref{fig:normal-embeddings-analytical-propertiesB}) with higher maximum degrees than seen in for uniform node embeddings. 

Figure \ref{fig:normal-embeddings-analytical-propertiesCD} shows that the density of the hard geometric MARS network with normal node embeddings is significantly higher than the density of the space-independent case or the hard geometric case with uniform node embeddings, which can be explained by the very high intra-layer degrees. Even when the overlap in edges across layers is high, having a high intra-layer degree increases density significantly because of the composition of cliques structure of each layer. Figure \ref{fig:normal-embeddings-analytical-propertiesCD} also shows that the clustering coefficient is similar in the case of normal node embeddings to the case of uniform node embeddings.

We explore this configuration of the MARS model in the context of real-world register-based networks through simulation in Section \ref{sec:experiments}.

%% file: section/6-experiments/experiments.tex
In this section we demonstrate how the MARS framework can be used to model real-world register-based social networks. First we discuss the experimental setup in Section \ref{subsec:trun-experimental-setup} and configure a small-scale MARS network ensemble in Section~\ref{subsec:fitting}. 
In Section~\ref{subsec:control-mobility}, we investigate the parameter space of the connection function to explore the effects of spatial freedom on network properties. 

\subsection{Experimental setup}
\input{section/6-experiments/trunc-experimental-setup}
\label{subsec:trun-experimental-setup}

\subsection{Fitting model parameters}
\input{section/6-experiments/fitting}
\label{subsec:fitting}

\subsection{Effects of spatial freedom on drivers of social cohesion}
\input{section/6-experiments/control-mobility}

\label{subsec:control-mobility}

%% file: section/6-experiments/trunc-experimental-setup.tex
We let $N = 10^4$, a number small enough to run experiments on yet large enough to avoid sampling noise in the properties we observe. We assume $L=5$ layers, as is common in register-based social networks studied across Europe \cite{netherlands-anatomy,danish-anatomy,swedish-anatomy}.
The metric space of a register-based social network is the bounded region of a country with two-dimensional coordinates (longitude and latitude). Accordingly we define $\mathbf S$ to be a bounded subset of $\sqR$, namely the unit square $[0, 1]^2$ studied in Section \ref{subsubsec:hard-geometric-case}. Then nodes and affiliations have two-dimensional embeddings and the metric $d$ is Euclidean distance. 
As is common in literature we assume that affiliations are distributed uniformly in $\mathbf S$, but to account for the diversity in affiliation sizes commonly observed in population-scale networks, we embed individual nodes according to a two-dimensional truncated normal distribution with zero covariance, i.e. 
\begin{align*}
    X_V, Y_V \stackrel{\text{i.i.d.}}{\sim} \text{TruncNormal}(0.5,\sigma^2, 0, 1).
\end{align*}
We expect that affiliations located near the centre of the space will attract more nodes than those at the periphery, due to the higher density of nodes in the centre. The parameter $\sigma$ controls the spread of nodes across $\mathbf S$; a high value of $\sigma$ implies a more even distribution of nodes across $\mathbf S$, and a small value of $\sigma$ implies a densely packed core with few peripheral nodes. We fit the parameter $\sigma$ to the real data in Section \ref{subsec:fitting}.

Van der \textcite{popnet-data} provides the number of components $\# components_\ell$ in each layer of the population-scale network of the Netherlands in 2018. Equating this number to the number of cliques that the MARS model generates in each layer, we choose the number of affiliations $K^\ell$ by scaling $\#components_\ell$ from 17,254,523 nodes to 10,000 nodes. The scaled layer sizes are given in Table \ref{tab:layers}, rounded to the nearest integer.

\begin{table}[ht]
\centering
\begin{tabular}{lrr}
\toprule
Layer  & $\#components_\ell$ & Layer size ($K$)\\
\midrule
Family & $1,132,025$ & 656 \\
Household & $8,108,998$ & 4700 \\
Neighbors & $563,324$ & 326 \\
School & $13,737,622$ & 7962 \\
Work & $10,036,529$ & 5817 \\
\bottomrule
\end{tabular}
\caption{Definition of the five network layers and the number of affiliations per layer obtained by scaling the number of components per layer in the population-scale network of the Netherlands \cite{netherlands-anatomy} to the sampled MARS network. 
} 
\label{tab:layers}
\end{table}

We apply the exponential decay connection function of the soft RGG model \cite{waxman} defined in Equation \ref{eq:weighted-distance-function}, where $r_0 = \sqrt 2$ in the unit square. Omitting the parameter $\beta$, which is redundant under normalisation, we obtain
\begin{equation*}
    \w{u}{\al{i}} = \exp\left({-\frac{d(u, \al{i})}{\alpha \cdot \sqrt 2}}\right).
\end{equation*}
In this context the parameter $\alpha$ represents \textit{spatial freedom}. As $\alpha$ increases, the significance of $d(\xu, \xal{i})$ in the exponent of $\omega$ decreases, meaning the choice of affiliation for each node depends less on their proximity. Conversely, for small values of $\alpha$, $d(\xu, \xal{i})$ dominates the exponent and the system becomes more spatially dependent. In our experiments, we aim to observe how the synthetic network's properties change as we vary $\alpha$.

%% file: section/6-experiments/fitting.tex
To find a MARS ensemble which reproduces the properties of the degree distribution of the real register-based network of the Netherlands \cite{fragmentation}, we must choose appropriate values for the spatial freedom parameter $\alpha$ and the standard deviation of the node embedding distribution $\sigma$, described in Section~\ref{subsec:trun-experimental-setup}. We fit the parameters by minimising the mean squared error (MSE) over summary statistics of the real degree distribution in 2018~\cite{fragmentation}. In particular we consider the mean, median, and  25\textsuperscript{th} and 75\textsuperscript{th} percentile of the degree distribution, which are 126.1, 97, 52 and 185 respectively \cite{fragmentation}.
Initial exploration of the parameter domain revealed that these properties are most closely replicated for small values of $\sigma$, implying a system with an uneven density of nodes. We therefore vary over values of $\sigma$ in the set $\{0.1, 0.125, 0.15, 0.175, 0.20\}$. We explore $\alpha$ over all values $2^{-\frac{i}{2}}$ for $i = 0, ..., 20$, to cover the spectrum of spatial freedom from the completely space-independent case (Section \ref{subsubsec:space-independent-case}) to the hard geometric case (Section \ref{subsubsec:hard-geometric-case}).

\begin{figure}[!t]
    \centering
    \begin{subfigure}[t]{0.55\linewidth}
        \centering
        \includegraphics[width=\linewidth]{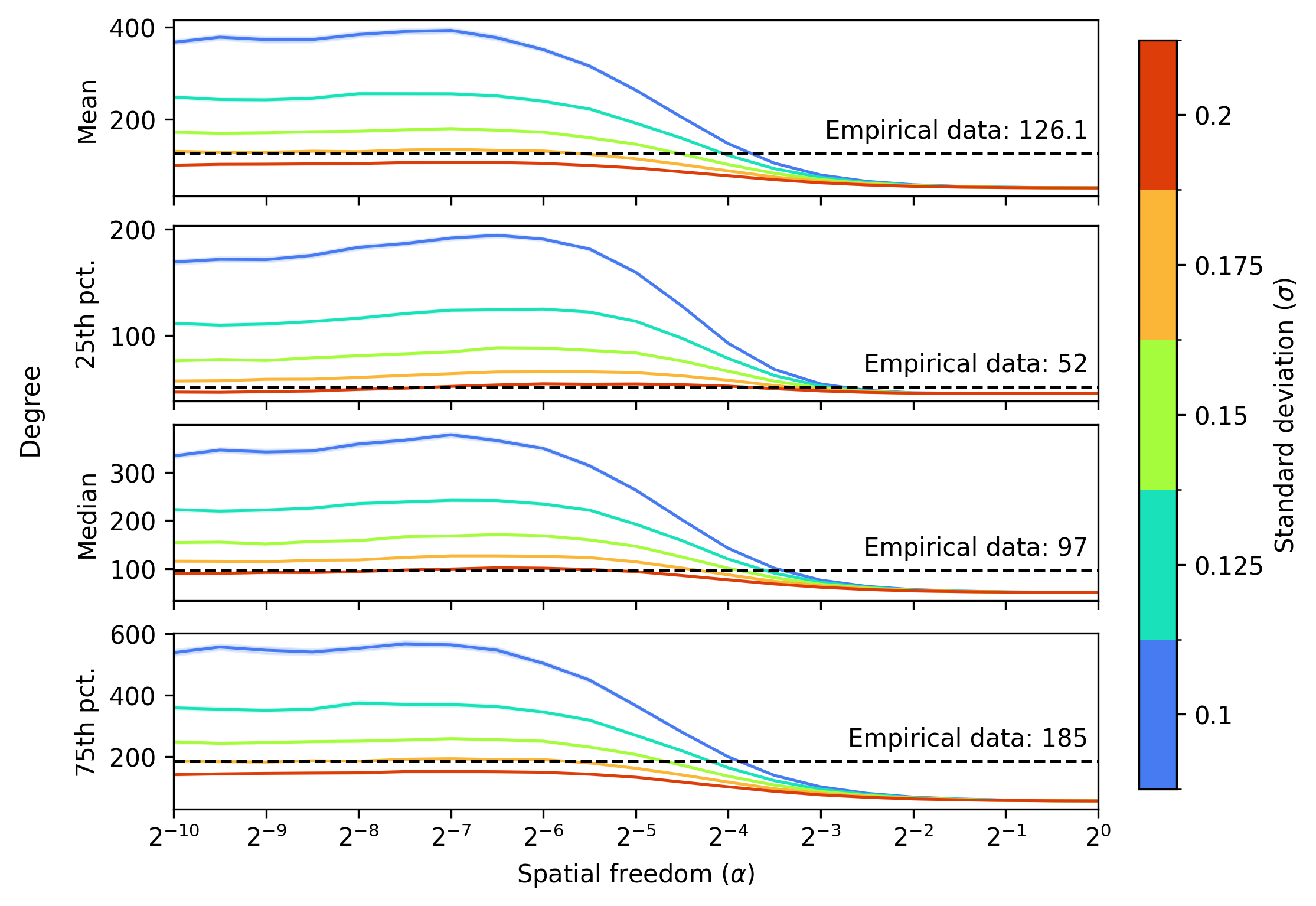}
        \caption{}
        \label{fig:deg-stats}
    \end{subfigure}
    \hspace{1.2em}
    \begin{subfigure}[t]{0.41\linewidth}
        \centering
        \includegraphics[width=\linewidth]{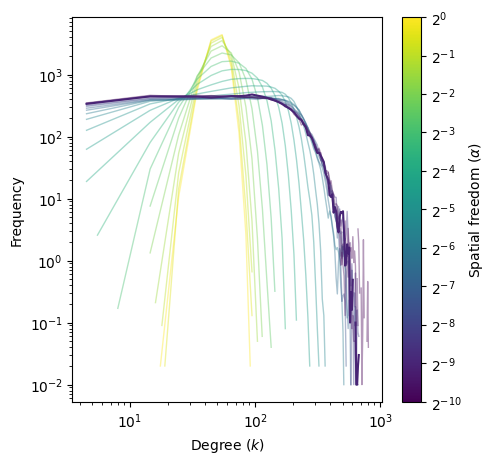}
        \caption{}
        \label{fig:deg-chosen}
    \end{subfigure}
    
    \caption{Comparison of the degree distribution for different values of $\sigma$ and $\alpha$. (a) Mean, 25\textsuperscript{th} percentile, median, and 75\textsuperscript{th} percentile of the degree distribution. 
    Dashed lines represent the empirical values of the degree distribution  for the register-based social network of the Netherlands in 2018~\cite{fragmentation}. MSE is minimised for $\alpha = 2^{-9}$ and $\sigma = 0.175$. 
    (b) Average degree distribution for varying $\alpha$ with $\sigma = 0.175$. The bold curve shows $\alpha = 2^{-9}$. The degree distribution transitions from a gamma distribution for small $\alpha$ (Section \ref{subsubsec:space-independent-case}) to a binomial distribution for large $\alpha$ (Section \ref{subsubsec:hard-geometric-case}).}
    \label{fig:mean-median-degree-fit}
\end{figure}

We generate 100 instances of a MARS network for each ($\sigma$, $\alpha$) pair. The summary statistics of their degree distributions are shown in Figure~\ref{fig:deg-stats}, where we compare the empirical values (indicated by the dotted lines) to the mean values of the MARS ensembles for all values of $\sigma$ and $\alpha$. 
We find that the MSE is minimised to $93.22$ when $\alpha = 2^{-9}$ and $\sigma = 0.175$, with a mean degree of $128.4$, a median degree of $114.7$, and 25\textsuperscript{th} and 75\textsuperscript{th} percentiles of $59.1$ and $182.7$. This implies a system with low spatial freedom where nodes strongly prefer nearby affiliations, and combined with a small $\sigma$ leads to an uneven distribution of affiliation sizes. The degree distributions resulting from all combinations of $\alpha$ values with $\sigma = 0.175$ are shown in Figure~\ref{fig:deg-chosen}, with the $\alpha = 2^{-9}$ curve in bold. As $\alpha$ increases, we see that the degree distribution transitions from a gamma distribution in the hard geometric case for low $\alpha$ (Section \ref{subsubsec:hard-geometric-case}) to a binomial distribution in the space-independent case for high $\alpha$ (Section \ref{subsubsec:space-independent-case}). In between the degree distribution fits no recognisable shape, but is similar to the degree distribution seen in the real population-scale network of the Netherlands \cite{netherlands-anatomy}. The summary statistics of the degree distributions obtained with all ($\sigma$, $\alpha$) pairs and their MSE with respect to the empirical properties are reported in Appendix \ref{app:fitting}.

\begin{figure}[!b]
    \centering
    \begin{subfigure}{0.38\linewidth}
        \centering
        \includegraphics[width=\linewidth]{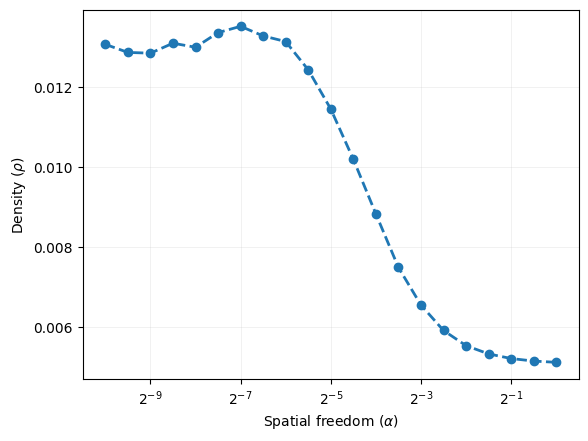}
        \caption{}
    \end{subfigure}
    \hspace{1.2em}
    \begin{subfigure}{0.38\linewidth}
        \centering
        \includegraphics[width=\linewidth]{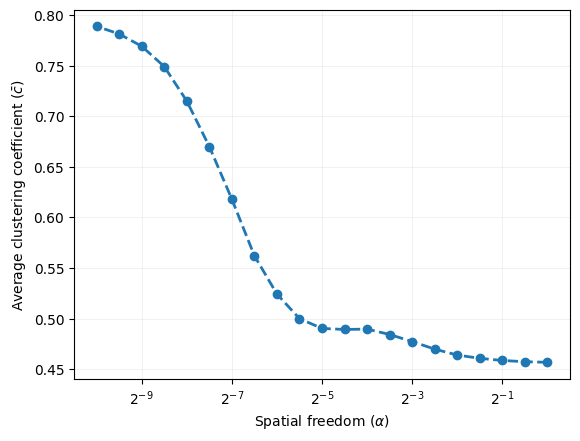}
        \caption{}
    \end{subfigure}
    
    \caption{Values of (a) the network density $\rho$ and (b) the average local clustering coefficient $\bar{c}$ for ensembles of MARS networks with the parameters in Section \ref{subsec:trun-experimental-setup}, $\sigma = 0.175$ and varying values of spatial freedom $\alpha$. 
    }
    \label{fig:basic-properties}
\end{figure}

Figure \ref{fig:basic-properties} shows how the density $\rho$ and average local clustering coefficient $\overline c$ varies with $\alpha$. As described in Section~\ref{subsec:limiting-cases}, we see that the density decreases as $\alpha$ increases due to the increased variation in affiliation sizes introduced by the normal node embeddings and hard geometric boundaries. When $\alpha$ is small $\rho$ converges to a value of approximately 0.013, and when $\alpha$ is high $\rho \approx 0.005$,  given by Equation ($\ref{eq:uniform-case-density}$). The average local clustering coefficient $\overline c$ also decreases as $\alpha$ increases, from approximately $0.79$ when $\alpha$ is small to $0.46$ when $\alpha$ is large, as given by Equation (\ref{eq:uniform-case-clustering}). We reasoned in Section \ref{subsec:limiting-cases} that this can be explained by increased rates of inter-layer triadic closure when spatial freedom is low. We note that these values of density and clustering are higher than expected for register-based social networks, when compared to the density $\rho = 0.000005$ and average local clustering coefficient $\overline c = 0.40$ of the network of the Netherlands \cite{netherlands-anatomy}.  This is likely due to the modelling of each layer of a MARS network as a composition of disconnected cliques, compared to the real data where connections are sparser either due to post-processing (e.g. limiting degrees in the work layer) or the construction method of the layer from register data (e.g. the family layer). We discuss this further and give recommendations for future work in Section \ref{sec:conclusion}. The complete list of density and average local clustering values for each value of $\alpha$ is reported in Appendix \ref{app:properties-table}.

For the remainder of our experiments, we fix the standard deviation $\sigma$ at $0.175$, and focus on the effect of varying the spatial freedom $\alpha$ on network properties to study how geographic mobility and the amount of overlap in contacts across social contexts affect global network properties.

%% file: section/6-experiments/control-mobility.tex
\textcite{fragmentation} observe a decline in social cohesion in the register-based network of the Netherlands between 2010 and 2021, and identify greater geographic mobility and decreasing overlap in social contexts (multiplexity) as the drivers of this change. To analyse this hypothesis using the MARS model, we will first show that the parameter $\alpha$ of our chosen connection method in Section \ref{subsec:trun-experimental-setup} controls mobility and multiplexity, and then use this to examine measures of social cohesion under different values of $\alpha$. 

\begin{figure}[!b]
    \centering
    \begin{subfigure}{0.4\linewidth}
        \centering
        \includegraphics[width=\linewidth]{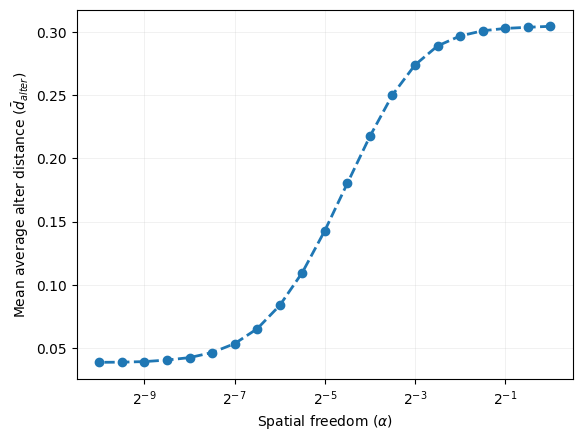}
        \caption{}
    \end{subfigure}
    \hspace{1.2em}
    \begin{subfigure}{0.4\linewidth}
        \centering
        \includegraphics[width=\linewidth]{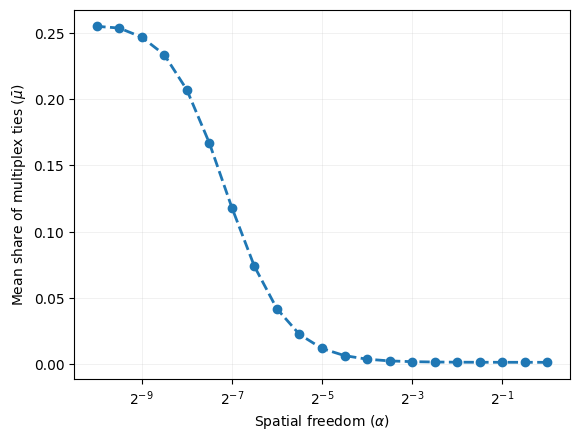}
        \caption{}
    \end{subfigure}
    
    \caption{The effect of spatial freedom $\alpha$ on (a) the mean average alter-to-alter distance $\bar{d}_{alter}$ and (b) the mean share of multiplex ties $\bar{\mu}$, representing mobility and overlap of social contexts, respectively. (a) $\bar{d}_{alter}$ is small ($\approx 0.04$) for low $\alpha$ and increases according to an s-curve to approximately 0.30 for high $\alpha$. (b) $\bar{\mu} \approx 0.25$ for low $\alpha$, and decreases according to an s-curve as $\alpha$ increases, converging to 0.00 when $\alpha \approx 2^{-4}$.}
    \label{fig:properties}
\end{figure}

Recall from Section \ref{subsec:register-based-social-networks} that mobility can be measured as the average alter distance (Equation~\ref{eq:alter-distance}) and multiplexity can be measured by a node's share of multiplex edges (Equation~\ref{eq:multiplexity}). A high average alter distance means that a node's connections are not geographically concentrated in a specific area, and thus implies greater mobility. A high share of multiplex edges means that a node frequently makes the same connections in different contexts, implying that the overlap in social contexts is high. Fixing $\sigma$ as $0.175$, we calculate the mean average alter distance and mean share of multiplex ties over all ego networks for all values of $\alpha$ and show the results in Figure \ref{fig:properties}. The average alter distance increases with $\alpha$ (Figure \ref{fig:properties}a), which is intuitive because for large $\alpha$ nodes are not limited by distance when choosing affiliations and thus connect to spatially diverse alters.  Figure \ref{fig:properties}b shows that the share of multiplex edges decreases as $\alpha$ increases, which happens because when there is low spatial freedom (small $\alpha$) nodes which are spatially close are likely to choose the same nearby affiliations in multiple layers. The exact values for the mean average alter distance and mean share of multiplex ties are reported for each $\alpha$ value in Appendix \ref{app:properties-table}. Using $\alpha$ to control for mobility and multiplexity, we can then explore how they drive social cohesion.

\begin{figure}[!t]
    \centering

\begin{tikzpicture}
\node[anchor=south west,inner sep=0] (img)
    {\includegraphics[width=0.8\linewidth]{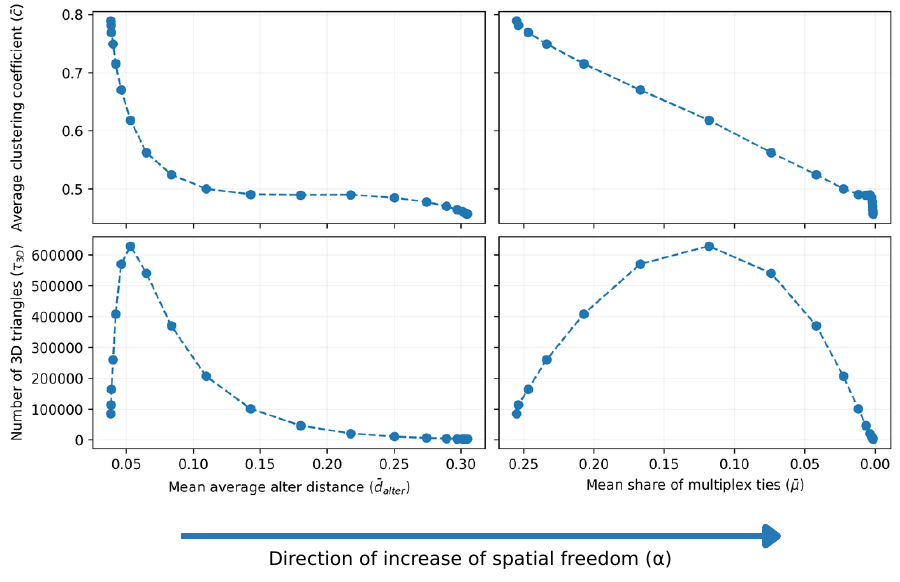}};

\node[anchor=north west,font=\bfseries]
    at ([xshift=-9mm,yshift=-17mm]img.north west) {(a)};

\node[anchor=north west,font=\bfseries]
    at ([xshift=-9mm,yshift=-50mm]img.north west) {(c)};

\node[anchor=north west,font=\bfseries]
    at ([xshift=145mm,yshift=-17mm]img.north west) {(b)};

\node[anchor=north west,font=\bfseries]
    at ([xshift=145mm,yshift=-50mm]img.north west) {(d)};

\end{tikzpicture}
    \caption{Measures of triadic closure against mean average alter distance $\bar{d}_{alter}$ and mean share of multiplex ties $\bar{\mu}$. The values of $\bar{\mu}$ are shown on a decreasing horizontal axis to keep the direction of increase of spatial freedom consistent. The average local clustering coefficient $\bar{c}$ decreases from $0.79$ to $0.46$ with (a) the increase of $\bar{d}_{alter}$ and (b) the decrease of $\bar{\mu}$. The number of three-dimensional triangles $\tau_{3D}$ shows a sevenfold increase from the initial value of $84,947$, but ultimately decreases to 2,824 as (c) $\bar{d}_{alter}$ increases and (d) $\bar{\mu}$ decreases.}
    \label{fig:cohesion}
\end{figure}
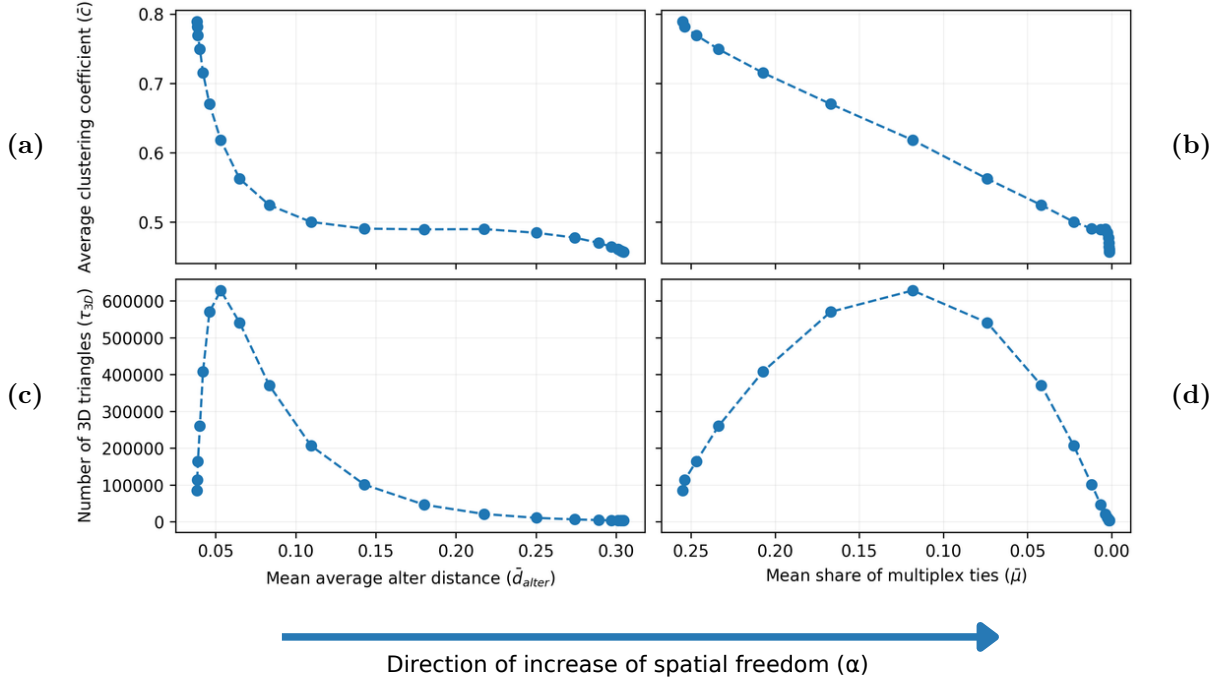

\textcite{fragmentation} measure social cohesion using the average local clustering coefficient (\ref{eq:local-clustering}), and conclude that greater geographic mobility and decreased overlap in social contexts leads to a decrease in cohesion. To investigate this we compare the average alter distance and share of multiplex edges, which we have shown to be controlled by $\alpha$, to the local clustering coefficient. In Figure \ref{fig:cohesion}, we see that as the mean average alter distance increases the average clustering coefficient decreases quickly, reaching a value of approximately 0.46 in the space independent case (Equation \ref{eq:uniform-case-clustering}). As the mean share of multiplex ties decreases we see that the average clustering coefficient also decreases, this time linearly. This is consistent with the findings of \textcite{fragmentation} and implies that greater mobility and decreased multiplexity indeed drive a decrease in social cohesion.

The MARS model introduces artificial closure through the composition of cliques structure of each layer, which is not caused by the underlying system. Therefore, to highlight triadic closure which is not caused by the network generative process, we also consider inter-layer closure, which we quantify by the number of three-dimensional triangles which do not have one-dimensional equivalents $\tau_{\text{3D}}$, as defined in Section \ref{subsec:monoplex-properties}. Figure \ref{fig:cohesion} shows  the relationship between $\tau_{\text{3D}}$ and the average alter distance and share of multiplex ties. In both cases we see a similar pattern: as the average alter distance increases or the share of multiplex edges decreases, $\tau_{\text{3D}}$ first increases and then, after a point, decreases. This suggests that modest increases in mobility or social context overlap facilitate the formation of triads spanning multiple contexts.

Initially, this increase is caused by the formation of more multilayer triangles. When average alter distance is very low or the share of multiplex edges is very high, nodes mostly choose the same alters in every layer and thus most multilayer triadic structures are also represented by one-dimensional equivalents. As we move away from the hard geometric case, nodes choose more diverse alters across layers and begin to form more unique three-dimensional triangles. However, as mobility continues to increase, nodes become too spatially dispersed for independent connections to converge and form triadic structures. We observe that the proportion of total triangles which are three-dimensional remains low in all cases, implying that measures of closure in the network are dominated by one-dimensional triangles introduced by the construction method. The average local clustering coefficient, number of three-dimensional triangles, average alter distance and share of multiplex ties are reported for each $\alpha$ value in Appendix \ref{app:properties-table}.

Through our experiments we see that high geographic mobility and low overlap in social contexts leads to low social cohesion, measured both by average local clustering coefficient and the number of three-dimensional triangles, in agreement with \textcite{fragmentation}. We also see that when geographic mobility is low and the overlap in social contexts is high, clustering measures are dominated by intra-layer triadic structures introduced by the construction method and the number of three-dimensional triangles remains low. This implies that the conclusions we can make about the social cohesion in register-based networks are strongly influenced by the construction methods of the network, and this should be considered when interpreting results on such networks.

%% file: section/7-conclusion/conclusion.tex
In this paper we have introduced the MARS framework for the modelling and analysis of register-based social networks. First we studied its statistical properties analytically, and showed that the degree distribution transitions from a binomial to a skewed distribution as edge formation becomes increasingly dependent on spatial proximity.

Secondly, we applied the MARS framework to modelling real-world data. In a case study of the real population-scale register-based network of the Netherlands we have shown that the MARS model reproduces properties of the network in the simple case where the metric space is a unit square, affiliations are embedded uniformly in the space and nodes are embedded normally around the centre. 
Our model best reproduces these properties when the standard deviation of the normal distribution is relatively small 
and spatial freedom is relatively low, 
implying a system where population density has a large spread and edge formation is highly spatially dependent. 

Thirdly and finally, we explored the effect of the model parameters on network properties, particularly focussing on spatial freedom.
We observed that increasing spatial freedom leads to a decrease in clustering coefficient. This can be explained by the fact that in systems with high spatial dependence triadic closure is common because of the transitive property of proximity (two people spatially close to me are close to each other). We also observe that as spatial freedom increases the share of multiplex triangles increases to a point and then decreases. We reason that this is because we transition from a system with low spatial freedom where triangles mainly form within layers to a system with high spatial freedom where triangles do not form even across layers, and in between reaches a balancing point where triadic closure is common but intra-layer triangles are not yet dominant. These observations agree with existing empirical findings \cite{fragmentation} that increasing mobility decreases social cohesion, and thus demonstrates the applicability of the model to real data.

There are various opportunities for extensions of this work. Whilst we have applied a simple version of the MARS model to the study of register-based networks, the real systems that this framework tries to capture have more complex metric spaces, embedding methods and connection functions. Additionally, the MARS model is built on the principles of spatial-embeddedness and a bipartite structure, but this only strictly holds true for a subset of the layers in register-based networks. Using more complex parameter sets or different models for different layers would bring the model closer to the real data.  

We saw that the MARS model produces very high values for density and clustering, even greater than those seen in register-based networks. An extension of this model would be a stochastic block model based MARS network, where nodes instead form pairwise connections based on assigned in- and out-group probabilities, where a node's group is determined by its affiliation. This would replicate the way in which some layers are degree-limited (i.e. work), and introducing non-zero out-group probabilities would replicate the almost bipartite structure of the neighbourhood and family layer. 

The model may also have applications in other contexts. In this article we have interpreted the metric space as geographical space, but it is also possible to interpret the embedding of nodes and affiliations as positions in some other dimensional space, or to include the addition of an extra dimension which represents an attribute such as node similarity. Under this interpretation, nodes are proximate in space to nodes with similar traits, and connect to affiliations which are aligned with these traits. This potentially opens up the possibility of applying MARS across a much broader range of different social network types commonly studied in the network science literature.

    

%% file: section/acknowledgments.tex

\subsection*{Funding information}
The authors acknowledge that they received funding in support for this research from the Dutch Research Council (NWO) through the SSHOC-NL project.

\subsection*{Code availability}
The Python implementation of the MARS model and the experiments reported in Section \ref{sec:experiments} are accessible at \url{https://github.com/IrinaMonaEpure/MARS}.

\subsection*{Author contributions}
KH and FWT conceived the study. KH developed the model and the statistical analysis. IE developed the code and led the analyses. All authors contributed to writing, reviewed the manuscript, and approved the
final version.

\subsection*{Competing interests}
The authors declare that they have no competing interests.

%% file: appendix/three-dimensional-triangles-space-independent-case.tex
Let $\triangle_{\text{1D}}(u,v,w)$ and $\triangle_{\text{3D}}(u,v,w)$ denote the events that a one-dimensional or three-dimensional triangle on nodes $(u,v,w)$ exists in $G$, respectively. To calculate $\overline \tau_{\text{3D}}$, the expected number of three-dimensional triangles in $G$ which do not have a one-dimensional equivalent, we need to find $\prob{\triangle_{\text{3D}}(u,v,w) \cap \neg \triangle_{\text{1D}}(u,v,w)}$.

For a set of three nodes $u$, $v$ and $w$ in layer $\ell$ (let $K := K^\ell$), there are five possible configurations. Either no edges are present between them with probability $\frac{(K-1)(K-2)}{K^2}$; one of the three possible edges is present, which happens with probability $\frac{K-1}{K^2}$; or all of the edges are present with probability $\frac{1}{K^2}$. For $u$, $v$ and $w$ to form a three-dimensional triangle in $G$ each of the edges must exist in at least one layer, but any one layer must not contain all three.

Assume that the one-dimensional triangle $(u, v, w)$ does not exist in any layer, and consider the subsequent cases where a three-dimensional triangle $(u, v, w)$ also does not exist in $G$. In this case, either $(u, v)$, $(u, w)$ or $(v, w)$ is not present in any layer, which happens for each edge with probability $\prod_{\ell = 1}^L \left(1 - \frac{K^\ell-1}{(K^\ell)^2} - \frac{1}{(K^\ell)^2} \right) = \prod_{\ell = 1}^L \left(1 - \frac{1}{K^\ell} \right)$. 
By the inclusion-exclusion principle we must also consider the cases where two or three of $(u, v)$, $(u, w)$ or $(v, w)$ are not present in the same layer. Any two of these edges are missing from the same layer with probability $\prod_{\ell=1}^L \left(1 - 2\frac{K^\ell - 1}{(K^\ell)^2} - \frac{1}{(K^\ell)^2}\right) = \prod_{\ell=1}^L \left(\frac{K^\ell - 1}{K^\ell} \right)^2$, and as above all three edges are missing from the same layer with probability $\prod_{\ell=1}^L \frac{(K^\ell -1)(K^\ell - 2)}{(K^\ell)^2}$. Thus, by the inclusion-exclusion principle, the probability that neither a one-dimensional nor three-dimensional triangle $(u, v, w)$ is present in $G$ is 
\begin{align*}
    \prob{\neg \triangle_{\text{3D}}(u,v,w) \cap \neg \triangle_{\text{1D}}(u,v,w)}
    =
    3\prod_{\ell=1}^L \left(1 -  \frac{1}{{K^\ell}}\right) 
    -
    3 \prod_{\ell=1}^L \left(\frac{K^\ell - 1}{K^\ell} \right)^2 
    +
    \prod_{\ell=1}^L \frac{(K^\ell -1)(K^\ell - 2)}{(K^\ell)^2},
\end{align*}
where the factors of 3 account for the number of ways to choose subsets of edges. By the law of total probability 
\begin{align*}
    \prob{\neg \triangle_{\text{1D}}(u,v,w)} = \prob{\triangle_{\text{3D}}(u,v,w) \cap \neg \triangle_{\text{1D}}(u,v,w)} + \prob{\neg \triangle_{\text{3D}}(u,v,w) \cap \neg \triangle_{\text{1D}}(u,v,w)}.
\end{align*}
From Section \ref{subsubsec:space-independent-case} the probability that one-dimensional triangle $(u, v, w)$ does not exist in any layer is
\begin{align*}
\prob{\neg \triangle_{\text{1D}}(u,v,w)} = \prod_{\ell = 1}^L \left(1 - \frac{1}{(K^\ell)^2} \right).
\end{align*}
Thus $\prob{\triangle_{\text{3D}}(u,v,w) \cap \neg \triangle_{\text{1D}}(u,v,w)}$ is given by
\begin{align*}
    \prod_{\ell = 1}^L \left(1 - \frac{1}{(K^\ell)^2} \right)
    -
    3\prod_{\ell=1}^L \left(1 - \frac{1}{{K^\ell}}\right) 
    +
    3 \prod_{\ell=1}^L \left(\frac{K^\ell - 1}{K^\ell} \right)^2  
    -
    \prod_{\ell=1}^L \frac{(K^\ell -1)(K^\ell - 2)}{(K^\ell)^2},
\end{align*}
and the expected number of three-dimensional triangles in the monoplex network $G$ is found by multiplying by the number of possible node triplets $N\choose 3$.

%% file: appendix/two-dimensional-wedges-space-independent-case.tex
Similarly, let $\wedge_{\text{1D}}(u, v, w)$ and $\wedge_{\text{2D}}(u, v, w)$ denote the events that a one- or two-dimensional wedge $(u,v,w)$ centred on $v$ exists in $G$, where by ``centred on $v$'' we mean that the wedge consists of edges $(u, v)$ and $(v, w)$. To calculate the expected number of two-dimensional wedges in $G$ which do not have one-dimensional equivalents $\overline \nu_{\text{2D}}$ we need $\prob{\wedge_{\text{2D}}(u,v,w) \cap \neg \wedge_{\text{1D}}(u,v,w)}$ for each ordered set of nodes $u$, $v$ and $w$. 

For the edges $(u, v)$ and $(v, w)$ in layer $\ell$ (let $K := K^\ell$), there are four possible configurations. Either neither edge is present with probability $\frac{(K-1)^2}{K^2}$; one of the two edges is present which happens for each edge with probability $\frac{K-1}{K^2}$; or both edges are present with probability $\frac{1}{K^2}$. As in the case for three-dimensional triangles, we will assume that the one-dimensional wedge $(u, v, w)$ does not exist in any layer, which happens with probability $\prod_{\ell = 1}^L \left(1 - \frac{1}{(K^\ell)^2} \right)$. 

Consider the cases where the two-dimensional wedge $(u, v, w)$ is also not present. Edge $(u, v)$ is not in any layer with probability $\prod_{\ell = 1}^L \left(1 - \frac{K^\ell-1}{(K^\ell)^2} - \frac{1}{(K^\ell)^2} \right) = \prod_{\ell = 1}^L \left(1 - \frac{1}{K^\ell} \right)$, and the same probability holds for edge $(v, w)$. Neither edge is in any layer with probability $\prod_{\ell = 1}^L \left (\frac{K^\ell - 1}{K^\ell} \right )^2$. Therefore, by inclusion-exclusion 
\begin{align*}
    \prob{\neg \wedge_{\text{2D}}(u,v,w) \cap \neg \wedge_{\text{1D}}(u,v,w)}
    = 2 \prod_{\ell = 1}^L \left(1 - \frac{1}{K^\ell} \right) - \prod_{\ell = 1}^L \left (\frac{K^\ell - 1}{K^\ell} \right )^2,
\end{align*}
and by the law of total probability
\begin{align*}
    \prob{\wedge_{\text{2D}}(u,v,w) \cap \neg \wedge_{\text{1D}}(u,v,w)}
    =
    \prod_{\ell = 1}^L \left(1 - \frac{1}{(K^\ell)^2} \right) 
    -
    2 \prod_{\ell = 1}^L \left(1 - \frac{1}{K^\ell} \right) 
    +
    \prod_{\ell = 1}^L \left (\frac{K^\ell - 1}{K^\ell} \right )^2.
\end{align*}
There are $N\choose 3$ possible node triplets, and there are 3 ways to arrange the nodes in a wedge. Therefore the expected number of two-dimensional wedges is $3\cdot {N \choose 3} \cdot \prob{\wedge_{\text{2D}}(u,v,w) \cap \neg \wedge_{\text{1D}}(u,v,w)}$.

%% file: appendix/fitting-tables.tex
\begin{longtable}{ccrrrrr}
    \caption{
        Degree statistics obtained for different values of the spatial freedom $\alpha$ and the standard deviation of the node embedding distribution $\sigma$. The MSE is calculated relative to the empirical mean, 25th percentile, median, and 75th percentile values of the degree distribution reported for the population-scale social network of the Netherlands in the year 2018 \cite{popnet-data}.
    }
    \label{tab:degree-statistics-fitting}\\
    \toprule
    Standard deviation ($\sigma$) &
    Spatial freedom ($\alpha$) &
    Mean &
    25th &
    Median &
    75th &
    MSE \\
    \midrule
    \endfirsthead
\multirow{11}{*}{$ 0.1 $} & $2^{-10}$ & 366.8 & 169.0 & 334.1 & 538.0 & 63,116.92 \\
 & $2^{-9}$ & 372.5 & 171.3 & 342.3 & 545.9 & 66,338.84 \\
 & $2^{-8}$ & 383.5 & 182.9 & 358.9 & 552.3 & 71,731.66 \\
 & $2^{-7}$ & 392.4 & 191.6 & 377.9 & 563.4 & 78,123.85 \\
 & $2^{-6}$ & 350.8 & 190.6 & 349.7 & 503.4 & 58,732.66 \\
 & $2^{-5}$ & 263.1 & 159.5 & 263.4 & 366.0 & 22,686.46 \\
 & $2^{-4}$ & 147.2 & 92.8 & 142.5 & 199.6 & 1,098.78 \\
 & $2^{-3}$ & 79.4 & 54.7 & 76.5 & 102.2 & 2,368.12 \\
 & $2^{-2}$ & 58.1 & 46.4 & 56.9 & 68.8 & 4,943.26 \\
 & $2^{-1}$ & 52.7 & 46.0 & 52.1 & 59.1 & 5,817.71 \\
 & $2^{0}$ & 51.4 & 46.0 & 51.0 & 56.6 & 6,055.36 \\
\midrule
\multirow{11}{*}{$ 0.125 $} & $2^{-10}$ & 247.9 & 111.6 & 222.9 & 358.8 & 16,115.88 \\
 & $2^{-9}$ & 242.1 & 110.9 & 222.0 & 350.7 & 15,001.86 \\
 & $2^{-8}$ & 255.3 & 116.4 & 235.3 & 374.5 & 18,974.45 \\
 & $2^{-7}$ & 255.1 & 123.9 & 242.0 & 369.3 & 19,200.12 \\
 & $2^{-6}$ & 239.0 & 125.0 & 234.5 & 345.0 & 15,645.10 \\
 & $2^{-5}$ & 191.5 & 113.5 & 192.5 & 269.4 & 6,081.28 \\
 & $2^{-4}$ & 121.9 & 78.6 & 120.1 & 164.1 & 423.46 \\
 & $2^{-3}$ & 74.3 & 52.7 & 72.2 & 94.4 & 2,876.88 \\
 & $2^{-2}$ & 57.2 & 46.1 & 56.3 & 67.3 & 5,071.40 \\
 & $2^{-1}$ & 52.6 & 46.0 & 52.0 & 58.9 & 5,841.75 \\
 & $2^{0}$ & 51.4 & 46.0 & 51.0 & 56.5 & 6,062.37 \\
\midrule
\multirow{11}{*}{$ 0.15 $} & $2^{-10}$ & 172.0 & 76.3 & 154.5 & 248.3 & 2,503.01 \\
 & $2^{-9}$ & 170.7 & 76.7 & 151.6 & 245.9 & 2,321.76 \\
 & $2^{-8}$ & 174.1 & 81.2 & 158.4 & 250.4 & 2,802.84 \\
 & $2^{-7}$ & 179.7 & 84.8 & 168.2 & 258.6 & 3,611.09 \\
 & $2^{-6}$ & 171.9 & 88.2 & 168.6 & 250.3 & 3,200.36 \\
 & $2^{-5}$ & 146.0 & 83.8 & 146.8 & 207.2 & 1,094.86 \\
 & $2^{-4}$ & 102.0 & 66.6 & 101.3 & 136.6 & 789.64 \\
 & $2^{-3}$ & 69.5 & 50.9 & 68.1 & 87.1 & 3,407.82 \\
 & $2^{-2}$ & 56.3 & 46.1 & 55.6 & 65.8 & 5,210.44 \\
 & $2^{-1}$ & 52.4 & 46.0 & 52.0 & 58.5 & 5,874.26 \\
 & $2^{0}$ & 51.3 & 46.0 & 51.0 & 56.4 & 6,073.38 \\
\midrule
\multirow{11}{*}{$ 0.175 $} & $2^{-10}$ & 130.6 & 57.4 & 115.9 & 185.4 & 101.43 \\
 & $2^{-9}$ & 128.4 & 59.1 & 114.7 & 182.7 & 93.22 \\
 & $2^{-8}$ & 129.8 & 60.8 & 118.5 & 185.3 & 138.00 \\
 & $2^{-7}$ & 135.1 & 64.3 & 127.0 & 193.5 & 300.86 \\
 & $2^{-6}$ & 131.3 & 66.1 & 126.2 & 190.6 & 276.77 \\
 & $2^{-5}$ & 114.4 & 65.3 & 114.5 & 162.5 & 280.65 \\
 & $2^{-4}$ & 88.3 & 58.2 & 87.7 & 117.7 & 1,522.01 \\
 & $2^{-3}$ & 65.6 & 49.4 & 64.8 & 81.0 & 3,880.54 \\
 & $2^{-2}$ & 55.4 & 46.1 & 54.9 & 64.2 & 5,349.80 \\
 & $2^{-1}$ & 52.2 & 46.0 & 52.0 & 58.2 & 5,900.53 \\
 & $2^{0}$ & 51.3 & 46.0 & 51.0 & 56.2 & 6,081.70 \\
\midrule
\newpage
\midrule
\multirow{11}{*}{$ 0.2 $} & $2^{-10}$ & 100.6 & 47.1 & 90.2 & 141.6 & 651.32 \\
 & $2^{-9}$ & 102.8 & 47.5 & 92.5 & 145.9 & 527.77 \\
 & $2^{-8}$ & 104.1 & 49.5 & 94.4 & 147.6 & 473.40 \\
 & $2^{-7}$ & 106.8 & 52.4 & 99.6 & 151.9 & 367.80 \\
 & $2^{-6}$ & 104.5 & 54.8 & 101.4 & 149.3 & 441.38 \\
 & $2^{-5}$ & 94.6 & 54.7 & 94.2 & 133.3 & 918.95 \\
 & $2^{-4}$ & 77.7 & 52.6 & 77.3 & 102.2 & 2,397.05 \\
 & $2^{-3}$ & 62.4 & 48.1 & 61.8 & 76.1 & 4,292.80 \\
 & $2^{-2}$ & 54.8 & 46.0 & 54.2 & 63.0 & 5,460.49 \\
 & $2^{-1}$ & 52.0 & 46.0 & 51.8 & 57.9 & 5,929.11 \\
 & $2^{0}$ & 51.3 & 46.0 & 51.0 & 56.3 & 6,080.30 \\
        \bottomrule
\end{longtable}

%% file: appendix/property-tables.tex
\begin{table}[H]
\centering
\caption{Network properties of the MARS model implemented in Section \ref{sec:experiments} for values of the spatial freedom parameter $\alpha$ equal to $2^{-\frac{i}{2}}$ for $i=0, ..., 20$, when the standard deviation of the node embeddings is $\sigma=0.175$.}
\label{tab:network_properties_by_alpha}
\begin{tabular}{lrrrrrr}
\toprule
\makecell{Spatial\\freedom\\($\alpha$)} & \makecell[t]{Density\\($\rho$)} & \makecell{Mean average\\alter distance\\($\bar{d}_{\mathrm{alter}}$)} & \makecell{Mean share of\\multiplex ties\\($\bar{\mu}$)} & \makecell{Average clustering\\coefficient\\($\bar{c}$)} & \makecell{Number of\\1D triangles\\($\tau_{1D}$)} & \makecell{Number of\\3D triangles\\($\tau_{3D}$)} \\
\midrule
$2^{-10}$ & 0.0131 & 0.0384 & 0.2549 & 0.7888 & $3.433 \times 10^{7}$ & $8.495 \times 10^{4}$ \\
$2^{-9.5}$ & 0.0129 & 0.0385 & 0.2536 & 0.7813 & $3.197 \times 10^{7}$ & $1.137 \times 10^{5}$ \\
$2^{-9}$ & 0.0128 & 0.0390 & 0.2468 & 0.7692 & $3.149 \times 10^{7}$ & $1.637 \times 10^{5}$ \\
$2^{-8.5}$ & 0.0131 & 0.0402 & 0.2336 & 0.7491 & $3.230 \times 10^{7}$ & $2.599 \times 10^{5}$ \\
$2^{-8}$ & 0.0130 & 0.0421 & 0.2071 & 0.7150 & $2.946 \times 10^{7}$ & $4.079 \times 10^{5}$ \\
$2^{-7.5}$ & 0.0134 & 0.0461 & 0.1670 & 0.6700 & $2.900 \times 10^{7}$ & $5.698 \times 10^{5}$ \\
$2^{-7}$ & 0.0135 & 0.0532 & 0.1182 & 0.6180 & $2.753 \times 10^{7}$ & $6.281 \times 10^{5}$ \\
$2^{-6.5}$ & 0.0133 & 0.0650 & 0.0738 & 0.5621 & $2.339 \times 10^{7}$ & $5.405 \times 10^{5}$ \\
$2^{-6}$ & 0.0131 & 0.0835 & 0.0418 & 0.5242 & $2.122 \times 10^{7}$ & $3.702 \times 10^{5}$ \\
$2^{-5.5}$ & 0.0124 & 0.1095 & 0.0225 & 0.4999 & $1.697 \times 10^{7}$ & $2.063 \times 10^{5}$ \\
$2^{-5}$ & 0.0114 & 0.1428 & 0.0118 & 0.4902 & $1.371 \times 10^{7}$ & $1.008 \times 10^{5}$ \\
$2^{-4.5}$ & 0.0102 & 0.1804 & 0.0064 & 0.4892 & $1.056 \times 10^{7}$ & $4.595 \times 10^{4}$ \\
$2^{-4}$ & 0.0088 & 0.2176 & 0.0038 & 0.4896 & $7.720 \times 10^{6}$ & $2.086 \times 10^{4}$ \\
$2^{-3.5}$ & 0.0075 & 0.2504 & 0.0025 & 0.4842 & $5.320 \times 10^{6}$ & $1.029 \times 10^{4}$ \\
$2^{-3}$ & 0.0066 & 0.2741 & 0.0019 & 0.4772 & $3.870 \times 10^{6}$ & $6.183 \times 10^{3}$ \\
$2^{-2.5}$ & 0.0059 & 0.2892 & 0.0016 & 0.4698 & $2.996 \times 10^{6}$ & $4.395 \times 10^{3}$ \\
$2^{-2}$ & 0.0055 & 0.2971 & 0.0015 & 0.4639 & $2.509 \times 10^{6}$ & $3.578 \times 10^{3}$ \\
$2^{-1.5}$ & 0.0053 & 0.3012 & 0.0014 & 0.4607 & $2.258 \times 10^{6}$ & $3.183 \times 10^{3}$ \\
$2^{-1}$ & 0.0052 & 0.3031 & 0.0014 & 0.4586 & $2.121 \times 10^{6}$ & $2.976 \times 10^{3}$ \\
$2^{-0.5}$ & 0.0052 & 0.3040 & 0.0014 & 0.4574 & $2.046 \times 10^{6}$ & $2.872 \times 10^{3}$ \\
$2^{0}$ & 0.0051 & 0.3047 & 0.0014 & 0.4567 & $2.008 \times 10^{6}$ & $2.824 \times 10^{3}$ \\
\bottomrule
\end{tabular}
\end{table}

%% file: references.bib
@article{SERNs,
   author = {L. Barnett and E. Di Paolo and S. Bullock},
   doi = {10.1103/PhysRevE.76.056115},
   pages = {056115},
   issue = {5},
   journal = {Physical Review E - Statistical, Nonlinear, and Soft Matter Physics},
   month = {11},
   title = {Spatially embedded random networks},
   volume = {76},
   year = {2007}
}

@article{MLSERNs,
   author = {Jürgen Hackl and Bryan T. Adey and Manilo De Domenico},
   doi = {10.1093/comnet/cny019},
   issue = {2},
   journal = {Journal of Complex Networks},
   month = {4},
   pages = {254-280},
   publisher = {Oxford University Press},
   title = {Modelling multi-layer spatially embedded random networks},
   volume = {7},
   year = {2019}
}

@article{netherlands-anatomy,
   author = {Eszter Bokányi and Eelke M. Heemskerk and Frank W. Takes},
   doi = {10.1038/s41598-023-36324-9},
   issue = {1},
   journal = {Scientific Reports},
   month = {12},
   pmid = {37280385},
   publisher = {Nature Research},
   title = {The anatomy of a population-scale social network},
   volume = {13},
   year = {2023}
}

@article{menyhert2025connectivity,
  title={Connectivity and community structure of online and register-based social networks},
  author={Menyh{\'e}rt, M{\'a}rton and Bok{\'a}nyi, Eszter and Corten, Rense and Heemskerk, Eelke M and Kazmina, Yuliia and Takes, Frank W},
  journal={EPJ Data Science},
  volume={14},
  number={1},
  pages={1--19},
  year={2025},
  publisher={Springer},
  doi={10.1140/epjds/s13688-025-00522-4s}
}

@article{takes2026population,
  title={Population-scale Social Network Analysis: Advances and Opportunities},
  author={Takes, Frank W},
  journal={Encyclopedia of Social Network Analysis and Mining},
  %volume={14},
  %number={1},
  %pages={1--19},
  year={2026},
  publisher={Springer}
}

@article{multilayer-networks,
   author = {Mikko Kivelä and Alex Arenas and Marc Barthelemy and James P. Gleeson and Yamir Moreno and Mason A. Porter},
   doi = {10.1093/comnet/cnu016},
   issn = {20511329},
   issue = {3},
   journal = {Journal of Complex Networks},
   month = {9},
   pages = {203-271},
   publisher = {Oxford University Press},
   title = {Multilayer networks},
   volume = {2},
   year = {2014}
}

@misc{multiplex-triadic-structures,
      title = {Triadic structures in multislice networks}, 
      author = {Kevin Ren and Tara Trauthwein and Gesine Reinert},
      year={2025},
      month = {4},
      eprint={2504.00508},
      archivePrefix={arXiv},
      primaryClass={math.PR},
      url = {http://arxiv.org/abs/2504.00508}, 
}

@article{waxman,
    author = {Bernard M Waxman},
    title = {Routing of Multipoint Connections.},
    journal = {IEEE Journal on Selected Areas in Communications},
    year = {1988},
    pages = {1617--1622},
    volume = {6},
    number = {9},
    doi = {10.1109/49.12889}
}

@article{second-factorial-moments,
    author = {Potts, RB},
    title = {Note on the Factorial Moments of Standard Distributions},
    journal = {Australian Journal of Physics},
    volume = {6},
    number = {4},
    pages = {498-499},
    year = {1953},
    month = {12},
    doi = {10.1071/PH530498}
}

@article{poisson-voronoi-distribution,
title = {On the size distribution of Poisson Voronoi cells},
journal = {Physica A: Statistical Mechanics and its Applications},
volume = {385},
number = {2},
pages = {518-526},
year = {2007},
doi = {10.1016/j.physa.2007.07.063},
author = {Járai-Szabó Ferenc and Zoltán Néda}
}

@article{social-network-review,
title = {Social Network Modelling},
author={Amati, Viviana and Lomi, Alessandro and Mira, Antonietta},
year = {2018},
journal = {Annual Review Statistics and Its Application},
volume = {5},
pages = {343 -- 369},
doi = {10.1146/annurev-statistics-031017-100746}
}

@article{popnet-data,
    author = {van der Laan, Jan and de Jonge, Edwin and Das, Marjolijn and Te Riele, Saskia and Emery, Tom},
    title = {A Whole Population Network and Its Application for the Social Sciences},
    journal = {European Sociological Review},
    volume = {39},
    number = {1},
    pages = {145-160},
    year = {2023},
    month = {02},
    issn = {0266-7215},
    doi = {10.1093/esr/jcac026}
}

@article{social-network-comparison,
title = {A comparative study of social network models: Network evolution models and nodal attribute models},
journal = {Social Networks},
volume = {31},
number = {4},
pages = {240-254},
year = {2009},
doi = {10.1016/j.socnet.2009.06.004},
author = {Riitta Toivonen and Lauri Kovanen and Mikko Kivelä and Jukka-Pekka Onnela and Jari Saramäki and Kimmo Kaski},
}

@article{swedish-anatomy,
    author = {Panayiotou, Georgios and Wohlert, Inga K. and Bask, Miia and Bask, Mikael and Magnani, Matteo and Mäkinen, Ilkka Henrik},
    title = {Anatomy of a Swedish population-scale network},
    journal = {Scientific Reports},
    year = {2025},
    volume = {15},
    issue = {1},
    doi = {10.1038/s41598-025-15966-x}
}

@article{danish-anatomy,
    author = {Cremers, Jolien and Kohler, Benjamin and Maier, Benjamin Frank and Eriksen, Stine Nymann and Einsiedler, Johanna and Christensen, Frederik Kølby and Lehmann, Sune and Lassen, David Dreyer and Mortensen, Laust Hvas and Bjerre-Nielsen, Andreas},
    title = {Unveiling the social fabric through a temporal, nation-scale social network and its characteristics},
    journal = {Scientific Reports},
    year = {2025},
    volume = {15},
    issue = {1},
    doi = {10.1038/s41598-025-98072-2}
}

@inproceedings{multilayer-RGG,
    title={Rainbow connectivity of multilayered random geometric graphs}, 
    ISBN={978-84-18979-38-5}, 
    booktitle={DMD 2024: Discrete Mathematics Days}, 
    publisher={Universidad de Alcalá},
    author={Josep Díaz and Öznur Yaşar Diner and Maria Serna and Oriol Serra},
    year={2024},
    pages={160–165} 
}

@article{AB-random-graph,
   author = {Clara Stegehuis and Lotte Weedage},
   doi = {10.1016/j.physa.2021.126460},
   journal = {Physica A: Statistical Mechanics and its Applications},
   month = {1},
   publisher = {Elsevier B.V.},
   title = {Degree distributions in AB random geometric graphs},
   volume = {586},
   year = {2022}
}

@article{voronoi-cell-size,
  author={ElSawy, Hesham and Sultan-Salem, Ahmed and Alouini, Mohamed-Slim and Win, Moe Z.},
  journal={IEEE Communications Surveys \& Tutorials}, 
  title={Modeling and Analysis of Cellular Networks Using Stochastic Geometry: A Tutorial}, 
  year={2017},
  volume={19},
  number={1},
  pages={167-203},
  doi={10.1109/COMST.2016.2624939}}

@article{voronoi-polygon,
 author = {A. Hayen and M. P. Quine},
 journal = {Advances in Applied Probability},
 number = {2},
 pages = {281--291},
 publisher = {Applied Probability Trust},
 title = {Areas of Components of a Voronoi Polygon in a Homogeneous Poisson Process in the Plane},
 volume = {34},
 year = {2002},
 doi={10.1239/aap/1025131218}
}

@misc{fragmentation,
      title={Fragmentation of a longitudinal population-scale social network: Decreasing structural social cohesion in the Netherlands}, 
      author={Eszter Bokányi and Yuliia Kazmina and Eelke M. Heemskerk and Frank W. Takes},
      year={2026},
      eprint={2602.00234},
      archivePrefix={arXiv},
      primaryClass={physics.soc-ph},
      url={https://arxiv.org/abs/2602.00234}, 
}

@article{yuliiasocnets,
title = {Socio-economic segregation in a population-scale social network},
journal = {Social Networks},
volume = {78},
pages = {279-291},
year = {2024},
issn = {0378-8733},
doi = {10.1016/j.socnet.2024.02.005},
author = {Yuliia Kazmina and Eelke M. Heemskerk and Eszter Bokányi and Frank W. Takes}
}

@misc{yuliia-scirep,
      title={Can social capital remedy structural inequality? Economic mobility in a longitudinal population-scale social network}, 
      author={Yuliia Kazmina and Eelke M. Heemskerk and Emilia van der Kooij and Eszter Bokányi and Frank W. Takes},
      year={2025},
      eprint={2508.05275},
      archivePrefix={arXiv},
      primaryClass={physics.soc-ph},
      url={https://arxiv.org/abs/2508.05275}, 
}

@article{javier,
    author = {Javier Garcia-Bernardo and Christine Hedde-von Westernhagen and Tom Emery and Albert Jan van Hoek},
    title = {Assessing COVID-19 transmission through school and family networks using population-level registry data from the Netherlands},
    journal = {Scientific Reports},
    year = 2024,
    doi = {10.1038/s41598-024-82646-7},
    issue = 1,
    volume = 14, 
    pages = {31248}
}

@unpublished{kieran-aer-insights,
    author = {Candogan, Ozan and König, Michael D. and Marray, Kieran and Takes, Frank},
    title = {Network Rewiring and Spatial Targeting: Optimal Disease Mitigation in Multilayer Social Networks},
    year = {2025},
    note = {Chicago Booth Research Paper No. 25-01},
    doi = {10.2139/ssrn.5106505}
}
